\documentclass[twocolumn,journal ]{IEEEtran}

\usepackage[T1]{fontenc}
\usepackage[cmex10]{amsmath}
\usepackage{amssymb}
\usepackage{amsthm}
\usepackage{acronym}
\usepackage{tabularx}
\usepackage{booktabs}
\usepackage{xcolor}
\usepackage{array}
\usepackage{threeparttable}
\usepackage{booktabs}
\usepackage[pdftex]{graphicx}
\usepackage{cite}
\usepackage{color}
\newtheorem{remark}{Remark}
\usepackage{comment}
\usepackage{pifont}
\newcommand{\cmark}{\ding{51}}
\newcommand{\xmark}{\ding{55}}
\newcommand{\pmark}{$\boldsymbol{\circ}$}

\usepackage{subcaption}
\usepackage{float}
\usepackage[font=small]{caption}
\usepackage[unicode=true,
 bookmarks=true,bookmarksnumbered=true,bookmarksopen=true,bookmarksopenlevel=1,
 breaklinks=false,pdfborder={0 0 0},pdfborderstyle={},backref=false,colorlinks=false]
 {hyperref}
\hypersetup{pdftitle={Your Title},
 pdfauthor={Your Name},
 pdfpagelayout=OneColumn, pdfnewwindow=true, pdfstartview=XYZ, plainpages=false}

\AtBeginDocument{%
  \setlength{\abovedisplayskip}{4pt}
  \setlength{\belowdisplayskip}{4pt}
  \setlength{\abovedisplayshortskip}{4pt}
  \setlength{\belowdisplayshortskip}{4pt}
}
\newcommand{\ii}{\imath}

\newcommand{\T}{^{\intercal}}

\newcommand{\bfg}[1]{\boldsymbol{#1}}

\acrodef{rl}[RL]{rate limiter}
\acrodef{der}[DER]{distributed energy resource}
\acrodef{dg}[DG]{distributed generation}
\acrodef{dso}[DSO]{distribution system operator}
\acrodef{ev}[EV]{electric vehicle}
\acrodef{pcc}[PCC]{point of common coupling}
\acrodef{pll}[PLL]{phase-locked loop}
\acrodef{qep}[QEP]{quadratic eigenvalue problem}
\acrodef{srf}[SRF]{synchronously rotating frame}
\acrodef{tso}[TSO]{transmission system operator}
\acrodef{vsi}[VSI]{voltage source inverter}
\acrodef{res}[RES]{renewable energy source}
\acrodef{ess}[ESS]{energy storage system}
\acrodef{lec}[LEC]{local energy community}
\acrodef{lem}[LEM]{local energy market}
\acrodef{dgu}[DGU]{distributed generation unit}
\acrodef{cpl}[CPL]{constant power load}
\acrodef{gfl}[GFL]{grid-following}
\acrodef{gfm}[GFM]{grid-forming}
\acrodef{ibr}[IBR]{inverter-based resource}
\acrodef{rsu}[RSU]{robust synchronization unit}
\acrodef{iot}[IoT]{Internet-of-Things}
\acrodef{scada}[SCADA]{supervisory control and data acquisition}
\acrodef{pmu}[PMU]{power measurement unit}
\acrodef{hmi}[HMI]{human machine interface}
\acrodef{fft}[FFT]{fast fourier transform}
\acrodef{chil}[CHIL]{controller hardware-in-the-loop}
\acrodef{qsg}[QSG]{quadratic signal generator}
\acrodef{facts}[FACTS]{flexible AC transmission system}
\usepackage{algorithm}
\usepackage{algpseudocode}
\makeatletter
\let\oldforeign@language\foreign@language
\DeclareRobustCommand{\foreign@language}[1]{%
  \lowercase{\oldforeign@language{#1}}}
\theoremstyle{plain}

\theoremstyle{remark}

\makeatother

\definecolor{blue1}{RGB}{33,92,175}

\title{Exposing the Invisible: Detecting Stealthy Parameter-Based Cyber-Attacks on Inverter Synchronization Loops} 
\author{Zaint A. Alexakis,~\IEEEmembership{IEEE Member}, Michal M. Drewniak, and Charalambos Konstantinou,~\IEEEmembership{IEEE Senior Member} \thanks{Z. A. Alexakis, M. M. Drewniak, and C. Konstantinou are with the Computer, Electrical \& Mathematical Sciences and Engineering (CEMSE) Division, King Abdullah University of Science and Technology (KAUST), Thuwal, 23955, Saudi Arabia. C.  Konstantinou is also with the Energy Institute, The University of Texas at Austin, Austin, Texas, 78712, United States. M. M. Drewniak is also with the Department of Power Electronics \& Energy Control Systems, AGH University of Krakow, Krakow 30-059, Poland.}
\vspace{-6mm}}

\begin{document}

\maketitle
\pagestyle{plain} 
\IEEEpeerreviewmaketitle

\begin{abstract}
%The increasing use of \ac{iot} technologies for the monitoring and control of \acp{ibr} has introduced cyber-physical vulnerabilities that can be exploited to degrade system performance and threaten power-system stability. Recent studies show that adversaries with access to \ac{hmi} systems can stealthily manipulate controller gains of grid-supporting resources, primarily targeting the \ac{pll} responsible for synchronization and transient performance of grid-following converters. Motivated by revised grid-code requirements mandating ancillary service provision, this paper presents a small-signal stability analysis that explicitly captures the dynamics of \ac{pll}-based frequency estimators used for droop control and evaluates the impact of adversarial gain tampering under weak-grid conditions. The results reveal a critical interaction between frequency estimation, control, and synchronization dynamics that can significantly reduce stability margins. To address the stealthy nature of such attacks, a modified \ac{pll} implementation is proposed that exposes gain variations through shifts in its equilibrium points while preserving the dynamic performance of the conventional \ac{srf} \ac{pll}. Experimental results validate the effectiveness of the proposed approach in detecting \ac{pll} cyber-attacks.

The increasing integration of \ac{iot} technologies for monitoring and control of \acp{ibr} has expanded the attack surface, enabling stealthy manipulation of controller parameters through vulnerable supervisory control interfaces. \Acp{pll} emerge as prime targets, as they interact with all control loops and critically influence the dynamic response of \ac{gfl} converters. This paper analyzes the underlying threat model to elucidate the mechanisms enabling such stealthy behavior and conducts a thorough stability and transient response analysis to characterize how \ac{pll} tampering can degrade system performance without necessarily destabilizing the system. The results reveal critical interactions among frequency estimation, control, and synchronization that can significantly reduce stability margins. To counter the stealthy nature of these attacks, a modified \ac{pll} is proposed that exposes gain variations through shifts in its equilibrium points while preserving conventional \ac{pll} performance. Experimental results validate the effectiveness of the proposed approach in detecting \ac{pll} cyber-attacks.

\end{abstract}
\vspace{-2mm}

\IEEEpeerreviewmaketitle{}
\vspace{-3mm}
\section{Introduction}
\vspace{-1mm}
The power grid is transitioning from centralized, generator-based operation to decentralized coordination of grid-connected distributed generation, including inverter-based-resources (IBRs) \cite{fan2023modeling}. While this shift supports decarbonization, it introduces challenges such as reduced system inertia and damping, formation of weak or isolated networks prone to voltage disturbances and instability, and increased cybersecurity risks \cite{9371238,10238347,survey_smart}. Poorly integrated Internet-of-Things (IoT) technologies in power plant coordination have expanded attack surfaces, allowing adversaries to exploit vulnerabilities in critical communication channels and potentially disrupt grid operations \cite{fdi,cb_attacks,surf2}.

\Ac{scada}-based attacks are a credible threat, as evidenced by previous incidents. In particular, the cyber-attack reported in \cite{report} demonstrates how adversaries can disrupt communication between utilities and \acp{ibr} in California and Wyoming, exposing critical vulnerabilities in operational infrastructure. Other prominent cases include the 2019 disruption of communications with multiple renewable installations in Utah, and recent attacks on Poland's electricity grid \cite{pol_report}, in which adversaries bypassed local firewalls and erased supervisory device firmware without forensic traces. Beyond disrupting communication and supervisory functions, vulnerable \ac{scada} interfaces can also be exploited to target grid stability directly. This is exemplified in \cite{rev_scada,rev_scada2}, where the control parameters of \ac{facts} devices are manipulated to alter the reactance of multiple transmission lines, compromising the grid's frequency stability.

In this context, \acp{pll} emerge as particularly vulnerable targets, as they strongly influence \ac{ibr} transient behavior through their interactions with multiple control loops \cite{bali,dq_pll}. This vulnerability is further exacerbated by revised grid codes such as IEEE 1547 \cite{IEEE1547-2018}, which now mandate that \ac{gfl} converters, previously focused on market participation and maximum power point tracking, implement primary control loops to support grid frequency and voltage, similar to \ac{gfm} \acp{ibr} \cite{vsync_pll}. Meeting these requirements relies on accurate frequency estimation, conventionally derived from \acp{pll} \cite{fll1}. As a result, \acp{pll} are no longer used strictly for synchronization but also support controllers spanning longer time-scales, such as droop control, thereby dictating the overall small-signal profile of \acp{ibr}. This strong coupling between \acp{pll} and the various control layers may be exploited by adversaries seeking to harm this critical infrastructure: as illustrated in \cite{Exploiting_Inherent}, attackers exploiting vulnerabilities in \ac{ibr} communication channels can gain access to supervisory interfaces and manipulate critical \ac{pll} gains to degrade system operation, destabilize the network, or unnecessarily trigger protection mechanisms.

The severity of such attacks lies in their stealthy nature. As shown in \cite{Exploiting_Inherent}, if \ac{pll} tampering occurs at steady state, the system may appear unaffected, with the consequences manifesting only during subsequent disturbances. However, that work does not analyze the underlying mechanisms that render these attacks stealthy and evasive, and thus difficult to detect, which warrant further investigation. The same threat model is examined in \cite{Cyb_attacks}, where adversaries nullify the integral gain of the synchronization \ac{pll} to induce power deviations, exploiting the ability of the resulting proportional-only \ac{pll} to track grid frequency while introducing a steady-state phase error that produces power regulation errors in current-controlled \ac{gfl} \acp{ibr}. However, this attack is conspicuous rather than stealthy, as the loss of \ac{pll} phase lock leaves an immediate detection trace and is readily flagged by protection mechanisms. Furthermore, power shifting through this threat model can only be realized in the absence of dedicated primary controllers, a highly unlikely scenario given the mandate of IEEE 1547 \cite{IEEE1547-2018}. To mitigate this vulnerability, the authors propose an alternative synchronization scheme that neutralizes the attack at the expense of degraded system performance. The authors further introduce an attack detection method; however, it is restricted to the case in which the integral gain is nullified and fails to detect any other \ac{pll} gain tampering pattern. Similarly, \cite{DSOGI} presents a \ac{pll}-less current control strategy that addresses the considered threat model but does not surpass the performance of conventional \ac{pll}-based implementations. These limitations establish a clear need for a systematic framework capable of rigorously analyzing this attack vector, quantifying its impact on system performance, and exposing otherwise stealthy intrusions, while providing non-disruptive protection for vulnerable power electronic interfaces. %Collectively, these limitations establish a clear need for a systematic framework capable of rigorously analyzing this attack vector, quantifying its impact on system performance, and exposing otherwise stealthy intrusions, while providing non-disruptive protection for vulnerable power electronic interfaces.

The stakes of this gap are high: without an explicit characterization of how \ac{pll} tampering propagates through the control layers, operators cannot gauge an attack's severity or the conditions under which its consequences surface, a pressing concern for grid-supporting \ac{gfl} \acp{ibr} that rely on \acp{pll} for primary-control implementation. In \cite{Exploiting_Inherent,Cyb_attacks}, simulation-based analyses of grid-supporting \acp{ibr} highlight the impact of estimator dynamics on system performance and fault recovery time. However, an explicit small-signal analysis incorporating these supplementary loops is still lacking, particularly under weak-grid conditions where the control loops interact with the \ac{pll} used for synchronization. These analyses are essential to fully understand these interactions and to ensure that \ac{ibr} infrastructures remain stable and resilient to poor transients and potential instabilities.

This paper pursues two primary objectives. First, it formulates and rigorously analyzes a mathematical threat model for stealthy \ac{pll} tampering, explicitly capturing the mechanism by which the attack evades detection and quantifying its impact on \ac{ibr} dynamics through comprehensive small- and large-signal analyses. Prior works study such attacks, but do not analyze the mechanism underlying their stealthiness. Second, an effective detection framework is developed to safeguard critical \ac{ibr} infrastructure without compromising nominal control performance, making it straightforward to integrate into existing \ac{gfl} plants. Specifically, this work proposes a novel \ac{rsu}-based \ac{pll} architecture that renders both proportional and integral gain manipulations observable through shifts in equilibrium points while preserving the small-signal behavior of conventional \acp{pll}. This overcomes the limitations of existing approaches, which either detect only the special case of integral-gain nullification, degrade performance, or eliminate the \ac{pll} altogether, as summarized in the comparison in Table~\ref{tab:comparison}. A further strength of the framework is its generality, as the same equilibrium-based principle can be extended naturally to a broad class of \ac{ibr} control architectures, exposing parameter changes wherever they occur. Its effectiveness is demonstrated through extensive simulation studies and validated via \ac{chil} experiments.

\begin{table}[t]
\caption{ Comparison with the existing literature.}
\vspace{-1mm}
\label{tab:comparison}
\centering
 
\scriptsize
\renewcommand{\arraystretch}{1.12}
\setlength{\tabcolsep}{2.2pt}
\begin{tabularx}{\columnwidth}{@{}>{\raggedright\arraybackslash}Xcccc@{}}
\toprule
{\small\textbf{Feature}} 
& \cite{Exploiting_Inherent} 
& \cite{Cyb_attacks} 
& \cite{DSOGI} 
& \textbf{This work} \\
\midrule
{\small Threat model and stealthiness analysis}
& \pmark & \pmark & \pmark & \cmark \\
{\small Small-/large-signal attack impact assessment}
& \pmark & \xmark & \pmark & \cmark \\
{\small \ac{pll} gain tampering detection}
& \xmark & \pmark & \pmark & \cmark \\
{\small Nominal \ac{pll} performance preserved}
& \cmark & \xmark & \xmark & \cmark \\
{\small Industrial/\ac{gfl} controller compatibility}
& \cmark & \xmark & \xmark & \cmark \\
\bottomrule
\end{tabularx}
\\[2pt]
{\small \cmark~fully addressed \quad \pmark~partially addressed \quad \xmark~not addressed}
\vspace{-5mm}
\end{table}

The remainder of this article is organized as follows. Section II presents a detailed modeling framework for \ac{ibr}-dominated power systems, while Section III introduces a grid-code-compliant \ac{gfl} control framework based on dynamic frequency estimation. The threat model is formulated in Section IV, and Section V provides both small- and large-signal analyses to examine the interactions among the frequency estimator, synchronization and control layers, and other grid-connected resources. The proposed detection scheme is presented in Section VI, and experimental results validating its effectiveness and applicability are reported in Section VII. Finally, Section VIII concludes the paper.

\vspace{-2mm}
\section{System Modeling } 
\vspace{-1mm}
A model-based approach is required to comprehensively analyze \ac{pll}-targeting cyber-attacks, as adversarial gain tampering ultimately seeks to alter the small-signal characteristics of \acp{ibr}. Such analyses enable the approximate quantification of the impact of these cyber-attacks and facilitate the formulation of the most detrimental attack models, which inherently supports the design of effective detection schemes. Accordingly, this section presents a detailed model of \ac{ibr}-dominated networks, comprising a diverse mixture of \ac{gfl} and \ac{gfm} resources and their associated control schemes.

\vspace{-2mm}
\subsection{Power Converters}
\vspace{-1mm}
Power converters are commonly represented using equivalent averaged models, in which they are formulated as ideal voltage sources connected to the \ac{pcc} through an RL filter and a shunt capacitor. This system can be expressed using the Park transformation, which maps system states into a stationary two-axis ($dq$) reference frame, thereby facilitating analysis and control design. The current dynamics in the $dq$ domain are given by \cite{YazdaniIravani2010}:
\begin{align}
    l_{i}\ii_{d,i}^{\prime L_{i}}&=-r_{i}\ii_{d,i}^{L_{i}}+\omega_{p,i}l_{i}\ii_{q,i}^{L_{i}}+v_{f,d,i}^{L_{i}}-v_{c,d,i}^{L_{i}},\\
    l_{i}\ii_{q,i}^{\prime L_{i}}&=-r_{i}\ii_{q,i}^{L_{i}}-\omega_{p,i}l_{i}\ii_{d,i}^{L_{i}}+v_{f,q,i}^{L_{i}}-v_{c,q,i}^{L_{i}},
\end{align}
    where $i\in \mathcal{C}$, with $\mathcal{C}$ being the set of power converters. The parameters $r_{i}$, $l_{i}$ denote the $i_\text{th}$ converter parasitic resistance and inductance of the filter, $\omega_{p,i}$ is the \ac{ibr} frequency, derived from \acp{pll} for \ac{gfl} converters and from the swing equation for \ac{gfm} ones, $\ii_{d,i}^{L_{i}}$, $\ii_{q,i}^{L_{i}}$ are the converter output currents, $v_{f,d,i}^{L_{i}}$, $v_{f,q,i}^{L_{i}}$ and $v_{c,d,i}^{L_{i}}$, $v_{c,q,i}^{L_{i}}$ are the $dq$- frame inverter and capacitor voltages. The superscript $(\cdot)^{L_{i}}$ denotes quantities expressed in the $i_\text{th}$ local reference frame.
The capacitor voltages of the power converters' filters are given by:
\begin{align}
    \!\!\!\!C_{i}v_{c,d,i}^{'L_{i}}&\!\!=\!\!-\!v_{c,d,i}^{L_{i}}/\!R_{i}\!+\!\omega_{p,i}C_{i}v_{c,q,i}^{L_{i}}\!+\!\!\!\!\sum_{i\in\Omega_{j}}\!\!\ii_{in,d,i}^{L_{i}}\!-\!\!\!\sum_{i\in \mathrm{F}_{j}}\ii_{out,d,i}^{L_{i}},\label{eq:vd}\\ 
    \!\!C_{i}v_{c,q,i}^{'L_{i}}&\!\!=\!\!-\!v_{c,q,i}^{L_{i}}/\!R_{i}\!-\!\omega_{p,i}C_{i}v_{c,d,i}^{L_{i}}\!+\!\!\!\sum_{i\in\Omega_{j}}\!\!\ii_{in,q,i}^{L_{i}}\!-\!\!\!\sum_{i\in \mathrm{F}_{j}}\ii_{out,q,i}^{L_{i}},\label{eq:vq}
\end{align}
with $C_{i}$, $R_{i}$ denoting the capacitance and resistive load, and $v_{c,d,i}^{L_{i}}$, $v_{c,q,i}^{L_{i}}$ the capacitor $dq$-voltages. The summed terms correspond to the line currents entering and leaving the capacitor. Specifically, $\Omega_{j}$ and $\mathrm{F}_{j}$ denote, respectively, the sets of indices associated with currents entering and leaving the capacitor.
%In the considered model, capacitors are implemented in the filters of power converters and by load busses.

The voltage and current equations are defined in the local $dq$ frame associated with each power converter. Specifically, each subsystem operates in its own $dq$ frame, which enables the implementation of its controller and requires all system states to be expressed in the corresponding local coordinates.

\vspace{-2mm}
\subsection{Line, Load and Network Dynamics}
\vspace{-1mm}
The power network is modeled using strictly resistive-inductive lines and complex loads. 
Contrary to \cite{Exploiting_Inherent},  this paper considers the full nonlinear models of both lines and loads in order to accurately capture their interactions and their impact on the stability of \acp{ibr}. 
The dynamics of the transmission lines are described by:
\begin{align}
    l_{l,i}\,\imath_{l,d,i}^{\prime G} &= -r_{l,i} \imath_{l,d,i}^{G} + \omega_{g} l_{l,i} \imath_{l,q,i}^{G} + v_{s,d,i}^{G} - v_{e,d,i}^{G}, \\
    l_{l,i}\,{\imath}_{l,q,i}^{\prime G} &= -r_{l,i} \imath_{l,q,i}^{G} - \omega_{g} l_{l,i} \imath_{l,d,i}^{G} + v_{s,q,i}^{G} - v_{e,q,i}^{G},
\end{align}
where $i \in \mathcal{L}$, with $\mathcal{L}$ denoting the set of transmission lines. 
Here, $\imath_{l,d,i}^{G}$ and $\imath_{l,q,i}^{G}$ denote the line currents in the $dq$ reference frame; $r_{l,i}$ and $l_{l,i}$ are the line resistance and inductance, respectively; and $v_{s,d,i}^{G}$, $v_{s,q,i}^{G}$ and $v_{e,d,i}^{G}$, $v_{e,q,i}^{G}$ represent the sending- and receiving-end voltages in the $dq$ frame. The superscript $(\cdot)^{G}$ denotes quantities expressed in the global reference frame. The global reference frame is a common rotating frame with angular frequency $\omega_{g}$ and is used to mathematically interconnect the local reference frames of the converters. At steady state, it represents the grid frequency. Inductive loads can be modeled in the same manner by assuming that the receiving-end voltage of the corresponding line is connected to ground. Converter-free load buses can also be modeled using \eqref{eq:vd}-\eqref{eq:vq}; however, in this case, all states must be formulated in the global reference frame.
\vspace{-2mm}
\subsection{Generalized Power System Model}
\vspace{-1mm}
The distinction between local and global reference frames enables the modeling of synchronization dynamics for individual subsystems, including \ac{gfl} and \ac{gfm} \acp{ibr}. However, it complicates stability analysis, as transitioning between reference frames requires coordinate transformations. Suppose that $x^{L_{i}} = [x_{d}^{L_{i}},\, x_{q}^{L_{i}}]$ and $x^{G} = [x_{d}^{G},\, x_{q}^{G}]$ represent a power-system state expressed in the local and global reference frames, respectively. These states can be written in phasor form as:
\begin{align}
    X^{G} &= \left(x_{d}^{G} + j x_{q}^{G}\right)e^{j\!\!\int \omega_{g}\,dt}, \\
    X^{L_{i}} &= \left(x_{d}^{L_{i}} + j x_{q}^{L_{i}}\right)e^{j\!\!\int \omega_{p,i}\,dt}.
\end{align}
%where $j=\sqrt{-1}$ denotes the imaginary unit.

This representation is consistent with the $\alpha\beta$ transformation; therefore, a rotation matrix, $R$ \cite{Kundur1994}, can be used to transition between reference frames. The same concept applies to the phasor representation, yielding:
\begin{align}
    X^{G} = X^{L_{i}} e^{j\int \left(\omega_{p,i}-\omega_{g}\right)dt},
\end{align}
or, equivalently:
\begin{align}
    X^{G} = \left(x_{d}^{L_{i}} + j x_{q}^{L_{i}}\right)e^{-j\Delta \theta},
    \label{eq:localtoglobal}
\end{align}
where $\Delta \theta$ denotes the phase difference between the local and global frames and satisfies:
\begin{align}
    \Delta \dot\theta_i =  \omega_{g} - \omega_{p,i}.
\end{align}
Similarly, the transformation from the global reference frame to the local reference frame is given by:
\begin{align}
    X^{L_{i}} = \left(x_{d}^{G} + j x_{q}^{G}\right)e^{j\Delta \theta}.
    \label{eq:globaltolocal}
\end{align}
At steady state, $\omega_{g} = \omega_{p,i}^{\mathrm{s}}$ must hold for all $i \in \mathcal{C}$, where the superscript $(\cdot)^{\mathrm{s}}$ denotes steady-state, to ensure that the $dq$ quantities expressed in the global reference frame remain constant. This requirement facilitates stability analysis, as the corresponding equilibrium point is a constant vector, provided that the power system model is stable. It is emphasized that even if the reference frequency of the global frame differs from that of the local frames at steady state, the local $dq$ frames constructed by the power converters will still reach consensus on a common system frequency as a result of primary control. In this study, the global reference frequency $\omega_{g}$ is derived from the local reference frame of one of the power converters.

\vspace{-2mm}
\section{Ancillary Services-Oriented IBR Controllers}
%\vspace{-1mm}
Naturally, \ac{gfm} converters provide ancillary services such as frequency and voltage support. However, \ac{gfl} \acp{ibr} mainly focus on maximum power injection. 
%An elaborate depiction of a \ac{gfm} controller that emulates a virtual synchronous generator is given in Fig.~\ref{fig:gfm_control} . The droop curves which are indirectly implemented in the controller dynamics practically provide the necessary grid-support services. 
On the other hand, conventional \ac{gfl} schemes are not designed explicitly for real-time frequency support and can occasionally weaken the grid instead of strengthening it.
%Note that in the current modern smart-grid landscape \ac{gfl} \acp{ibr} significantly dominate installations of \ac{gfm} ones, thereby making the transition towards fossil fuel-free power networks even more challenging due to the potential reduced stability margins. For this reason, network operators have introduced standards and regulations that require \ac{gfl} implementations to offer both steady-state and real-time frequency support. 
%
\begin{figure}[t!]
    \centering
    \includegraphics[width=0.98\linewidth]{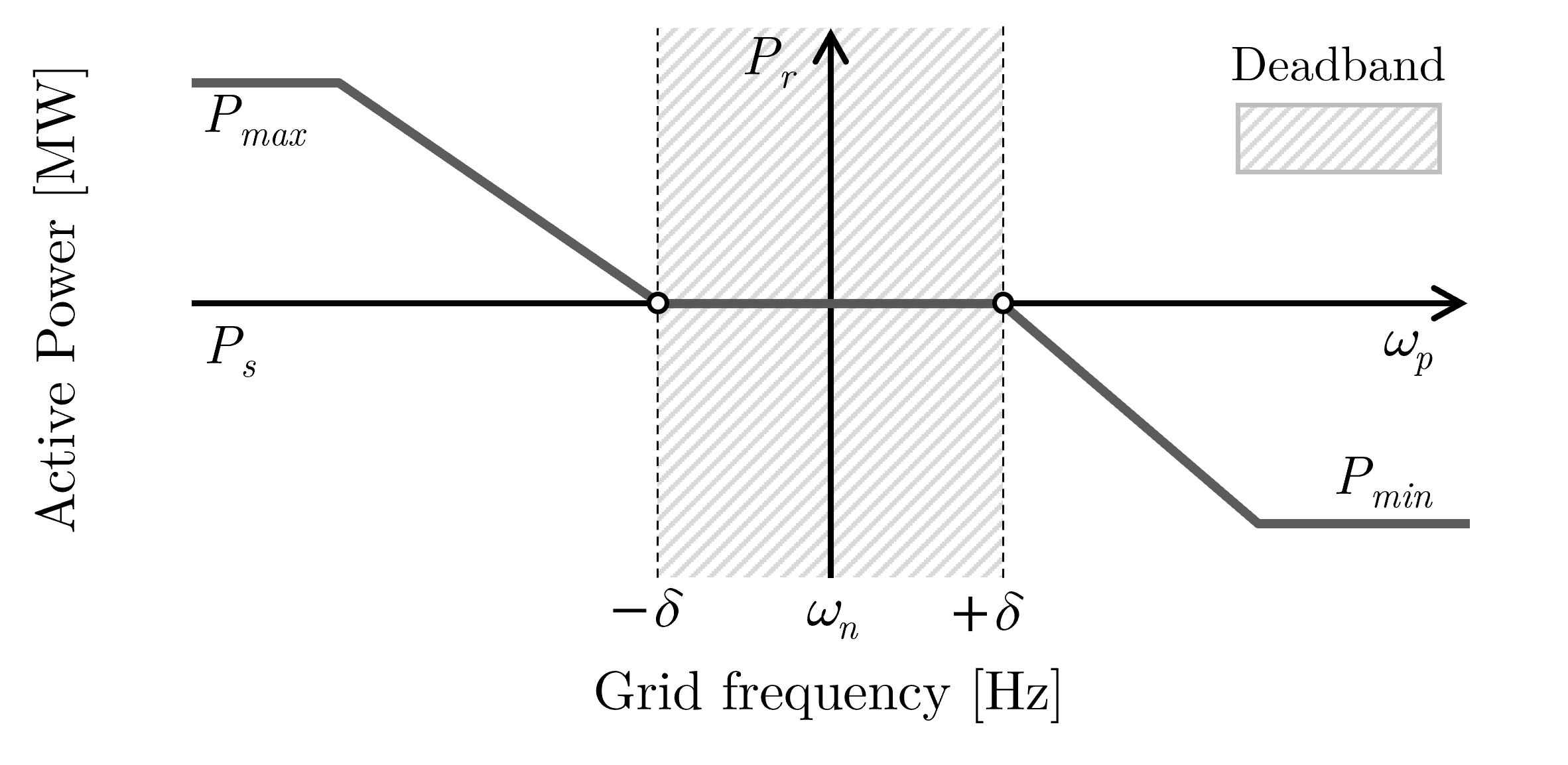}
    \vspace{-2mm}
    \caption{Frequency-support curve.}
    \label{fig:freq-sup}
    \vspace{-7mm}
\end{figure}
To address this concern, revised grid regulations, \cite{IEEE1547-2018}, require \ac{gfl} converters to strictly follow the droop-like characteristic illustrated in Fig.~\ref{fig:freq-sup}, where the power output of the converter is adjusted according to the system's frequency. These curves exhibit two key features: the \textit{deadband} and the \textit{slope}, the latter being equivalent to the active-power droop coefficient. Similarly, an analogous droop characteristic is specified for the reactive power control loop with respect to the \ac{pcc} voltage; however, it is omitted here to avoid redundancy. For clarity, the converter index $i$ is suppresed hereafter, as the following analysis focuses on a single representative \ac{gfl} converter, with the results applying identically to every $i\in\mathcal{C}$. The mathematical representation of the grid-support controller is given below:
\begin{align}
    P_{r}&=P_{s}+f\left(m_{P}\left(\omega_{n}-\omega_{g}\right)\right), \label{eq:p_ref} \\
    Q_{r}&=Q_{s}+f\left(m_{Q}\left(|v|_{n}-|v|_{pcc}\right)\right), \label{eq:q_ref}
\end{align}
where $\omega_{g}$ and $\omega_{n}$ correspond to the measured and nominal grid frequencies, $v_n$ and $v_{pcc}$ the nominal and measured \ac{pcc} voltage. Finally, $m_{P}$, $m_{Q}$ correspond to the active and reactive power droop gains, while $f$ denotes the deadband function:
\begin{equation}
f(x) =
\begin{cases} 
x - \delta, & \text{if } x > \delta,\\[0mm]
0, & \text{if } |x| \le \delta,\\[0mm]
x + \delta, & \text{if } x < -\delta,
\end{cases}
\end{equation}
where $\delta$ represents the deadband threshold. According to the ENTSO-E Frequency Sensitive Mode grid code \cite{ENTSOE2018_FSM}, the maximum value for $\delta$ is $0.5$~Hz. For $\delta=0$, the deadband is always inactive. 

\begin{figure}[t]
    \centering
    \includegraphics[width=0.95\linewidth]{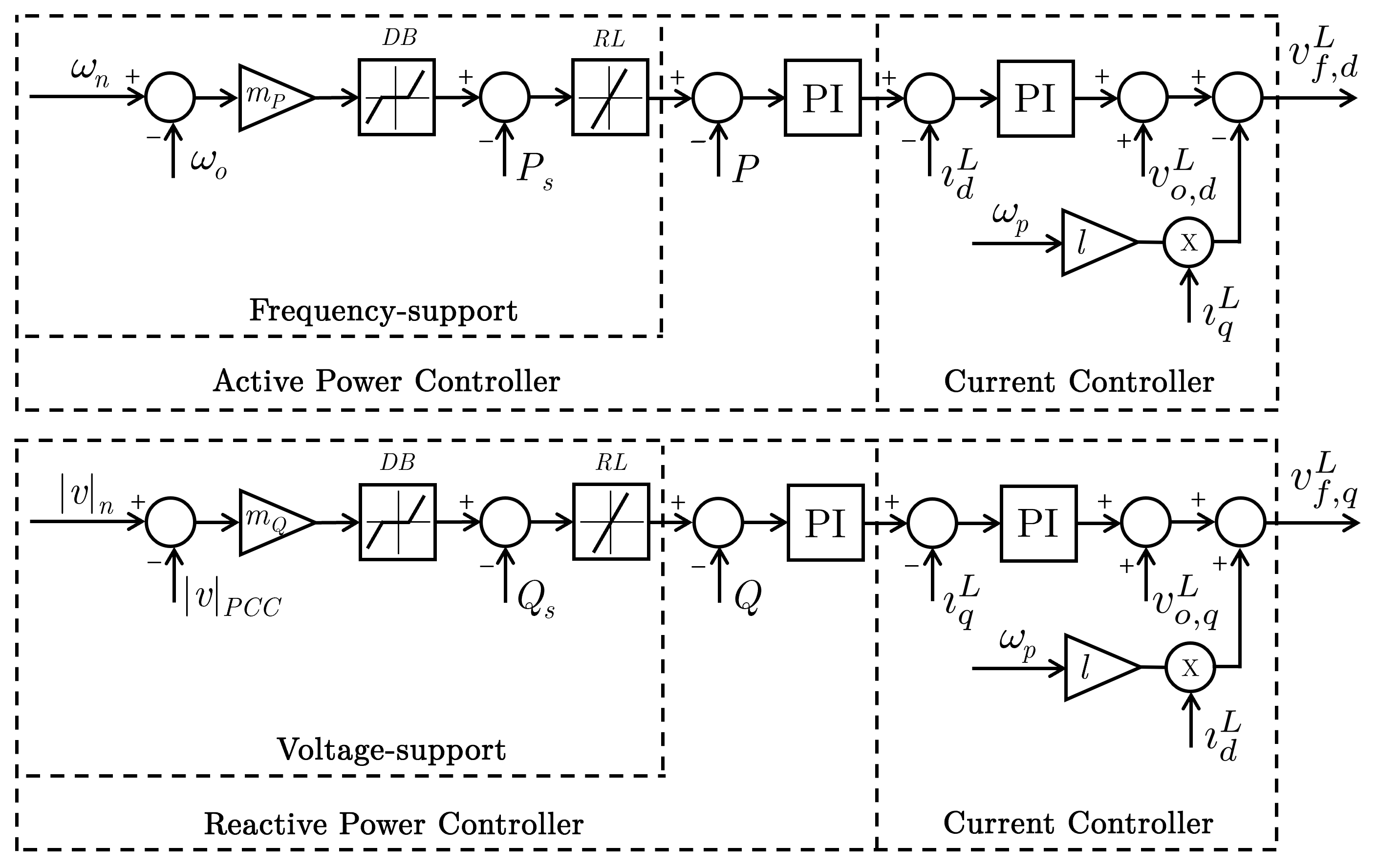}
    \vspace{-1mm}
    \caption{Grid-supporting \ac{gfl} control scheme.}
    \label{fig:gfl_control}
    \vspace{-6mm}
\end{figure}

The reference power is then usually channeled through a \ac{rl} to ensure that it adheres to physical plant restrictions and grid-codes: 
\begin{align}
    \dot P_{sr}=\dot P_{m}\text{sat}\left(a\left(P_{r}-P_{sr}\right)\right), \label{eq:p_rl}
\end{align}
where 'sat' denotes the saturation function, $\dot P_{m}$ the maximum allowed derivative limit and $a$ is a tuning parameter. The overall control diagram is illustrated in Fig. \ref{fig:gfl_control}.

%The basic notion of this controller is to provide active power support whenever the grid-frequency rises above or falls under the limits of the dead-zone. Inheretely, this structure follows the dynamics of droop-controlled \ac{gfm} converters, making the considered implementation a hybrid \ac{gfl} and \ac{gfm} control scheme.

%\begin{figure}[t!]
%    \centering
%    \includegraphics[width=1.0\linewidth]{figs/Control_system_GFM.png}
%    \caption{Grid Forming Inverter control system}
%    \label{fig:gfm_control}
%\end{figure}

The controller defined by \eqref{eq:p_ref} and \eqref{eq:p_rl} requires two inputs: the reference power $P_{s}$ and an estimate of the grid frequency. The method used to obtain $\omega_{g}$ significantly affects the converter's response during frequency support. \ac{gfl} converters implement \acp{pll} for both synchronization and frequency estimation for grid-monitoring purposes. However, \acp{pll} are tuned to act instantaneously, enabling the implementation of upper control layers. Consequently, the slightest grid-side disturbance significantly excites the \ac{pll} dynamics and its output frequency, which can drive an excessive reaction of the ancillary-service power controller despite the implemented deadband safeguard. This can be addressed by separating the synchronization and frequency-estimation functions into two distinct \acp{pll} that are tuned individually, so that a slower frequency-estimation \ac{pll} can be assigned independently of the synchronization dynamics. The same flexibility applies in the opposite case: under weak-grid conditions, where the synchronization \ac{pll} must be slowed to preserve stability, the separated structure still permits a responsive frequency-estimation \ac{pll}, decoupling the frequency-support response from the synchronization bandwidth.

This representation is made without loss of generality, since it is structurally equivalent to a single synchronization \ac{pll} augmented with additional frequency-estimation loops. Expressing the scheme as two \acp{pll} additionally allows each function, synchronization and frequency estimation, to be described and tuned in common \ac{pll} terms, such as bandwidth and damping ratio, so that the two loops are decoupled and the transient behavior and stability margins of each are inferred directly from well-established \ac{pll} design principles, as substantiated by the formal stability analysis %presented 
in Section~V.

Specifically, the synchronization layer implements the \ac{srf} \ac{pll} to track the phase, frequency and voltage magnitude of the grid:
\begin{align}    \omega_{p}&=\omega_{n}+k_{p,s}v_{c,q}^{L}+z,\label{eq:theta}\\
    \dot z&=k_{i,s}v_{c,q}^{L}, \label{eq:zeta}
\end{align}

\noindent with $k_{p,s}$ and $k_{i,s}$ being the proportional and integral gains of the \ac{pll}, and $\theta_{p}=\int\omega_{p}\,dt$ is the estimated input voltage phase. Since the analysis applies identically to both \acp{pll}, the subscript $s$ is omitted hereafter. The local-frame representation of the $q$-axis converter output voltage, $v_{c,q}^{L}$, can then be formulated using the transformation \eqref{eq:globaltolocal}:
\begin{align}
    v_{c,d}^{L}+jv_{c,q}^{L}=\left(v_{c,d}^{G}+jv_{c,q}^{G}\right)e^{j\Delta \theta},
\end{align}
where the auxiliary state $\Delta \theta$, representing the phase error between the global and local frames, is given by:
\begin{align}
    \Delta \dot \theta=\omega_{g}-\omega_{p}. \label{eq:tran}
\end{align}
Equating the real and imaginary parts yields:
\begin{align}
    v_{c,q}^{L}=v_{c,q}^{G}\cos\Delta \theta+v_{c,d}^{G}\sin\Delta\theta.
\end{align}
The outer-loop controller employs the same \ac{pll} formulation to estimate the grid frequency:
\begin{align}
    \omega_{o} &= \omega_{n} + k_{p,o} v_{c,q}^{Lo} + z_{o}, \label{eq:theta} \\
    \dot{z}_{o} &= k_{i,o} v_{c,q}^{Lo}, \label{eq:zeta}
\end{align}
where $k_{p,o}$ and $k_{i,o}$ denote the proportional and integral gains of the frequency estimator, respectively. Here, $\omega_o$ is the estimated frequency. The superscript $(\cdot)^{Lo}$ indicates quantities expressed in the reference frame of the outer-loop controller.

Following the same procedure used for the synchronization \ac{pll}, the outer-loop representation of the \ac{pcc} voltage is obtained by transforming it from the local \ac{pll} reference frame to the outer-loop reference frame:
\begin{align}
    v_{c,d}^{Lo} + j v_{c,q}^{Lo}
    = \left( v_{c,d}^{L} + j v_{c,q}^{L} \right) e^{j \Delta \theta_o}.
\end{align}
Here, $\Delta \theta_o$, similarly to $\Delta \theta$, is an auxiliary variable that represents the phase difference between the local reference frame and the outer-loop reference frame and is defined by:
\begin{align}
    {\Delta \dot \theta_o} = \omega_{p} - \omega_{o}.
\end{align}
%
%By using these two separate \acp{pll} for grid-synchronization and frequency estimation, the plant operator has now more leverage on navigating the plant behavior during events and under nominal conditions. For instance, by selecting a slower \ac{pll} response for the outer-loop \ac{pll} it ensures that the plant reacts to actual frequency disturbances and is not activated prematurely due to extreme \ac{pll} bandwidth. Thus, the synchronization \ac{pll} can be tuned to accommodate the operating conditions; if the system is connected to a weaker grid the operator can adapt the bandwidth to ensure system stability.  

Incorporating the ancillary services controller fundamentally alters the core operating principles of \ac{gfl} converters. In this context, a critical role is played by the frequency estimator, and its interaction with the synchronization \ac{pll}, as well as with other grid components such as \ac{gfm} converters, \acp{rl} and the deadband threshold. 
%Although stand-alone \acp{pll} have been extensively studied, a significant research gap remains regarding their integration within the proposed ancillary services scheme, particularly in the presence of the frequency estimator. 
These interdependencies may be exploited by adversaries seeking to degrade or potentially destabilize power networks by manipulating \ac{pll} gains that are accessible through supervisory control interfaces. To fully assess the potential impact of such malicious actions, a comprehensive threat model must be formulated to characterize both the nature and the consequences of the cyber-attack.
%, while a small-signal analysis of grid-connected \ac{gfl} converters equipped with the proposed ancillary services controller must be conducted in a networked environment to anticipate strategic gain manipulations by adversaries.
%\newpage
\vspace{-2mm}
\section{Attack Model Formulation}
\vspace{-1mm}
Building on the vulnerability exposed in the previous section, this section formalizes the attacker's objective and the notion of stealthiness underlying \ac{pll} gain-tampering attacks. Even when the frequency-estimation and synchronization \acp{pll} are properly tuned offline, both remain exposed to deliberate manipulation by adversaries who compromise supervisory control interfaces \cite{cb_attacks}, overwriting the proportional and integral gains of either loop. These interfaces, typically \acp{scada} and PLCs, can reportedly be breached by exploiting software vulnerabilities to obtain write access, whether through low-privilege accounts or remotely without prior system access, and to overwrite critical control parameters \cite{plc_scada,surf2}. The severity of this attack lies in its stealthy nature: altering the gains once the system has reached steady state produces no observable anomaly in the nominal response of the \acp{pll}, and the impact surfaces only when a subsequent disturbance excites the modified dynamics. The attack is practically manifested as a one-time parameter overwrite that requires no real-time observation by the adversary, a threat model also considered in~\cite{DSOGI,Exploiting_Inherent,Cyb_attacks}, and examined next to identify the factors that allow it to remain undetected.

Before proceeding with the analysis, the \ac{pll} gains prior to the attack are denoted as $k_{p}^{n}$ and $k_{i}^{n}$, and the gains after the attack as $k_{p}^{a}$ and $k_{i}^{a}$. The same analysis applies to both the frequency estimator and the synchronization \ac{pll}; hence, the gain subscripts are omitted, and the analysis is conducted for a generic \ac{pll} model. Accordingly, the objective of the attacker can be summarized as follows:
\begin{align}
    k_{p}^{n}\rightarrow k_{p}^{a},\quad k_{i}^{n}\rightarrow k_{i}^{a}. \label{eq:attack}
\end{align}
To analyze how an attack is perceived from the operator's perspective, the closed-loop \ac{pll} dynamics are reformulated by incorporating the coordinate transformation in \eqref{eq:tran} and assuming interconnection to an arbitrary grid voltage. Suppose that the operator has access to $v_{c,q}^{Lo}$ and the frequency estimate provided by the \ac{pll}. For $\hat{V} = 1$~pu, the $q$-axis voltage in the local frame is given by $v_{c,q}^{L} = \hat{V} \sin \Delta \theta$, yielding:
\begin{align}
    \Delta \dot \theta &= \omega_{g} - \omega_{n} - k_{p}^{n} \sin \Delta \theta - z, \\
    \dot z &= k_{i}^{n} \sin \Delta \theta.
\end{align}
The equilibrium points are yielded from:
\begin{equation}
    \begin{aligned}
        \Delta \theta^{\rm s}&=0,\quad z^{\rm s}=\omega_{g}-\omega_{n}. \label{eq:srfequil}
    \end{aligned}
\end{equation}
An attack is \emph{stealthy} if operators or monitoring systems that check the states against their expected steady-state values cannot detect it. Whether the adopted threat model is stealthy thus depends on its effect on the equilibrium. Let $\eta^{s}=\left(\Delta \theta^{\rm s},z^{\rm s}\right)^{n}-\left(\Delta \theta^{\rm s},z^{\rm s}\right)^{a}$ be the difference between the pre- and post-attack equilibria, where $\left(\Delta \theta^{\rm s},z^{\rm s}\right)^{n}$ and $\left(\Delta \theta^{\rm s},z^{\rm s}\right)^{a}$ denote the equilibrium pairs before and after the attack, as monitored by the operator. From \eqref{eq:srfequil}, tampering that preserves integral action leaves the equilibrium unchanged, so $\eta^{s}=0$, in agreement with the observations in \cite{Cyb_attacks,Exploiting_Inherent}. Consequently, if the system is at steady state when \eqref{eq:attack} occurs, it remains at equilibrium, unperturbed and without any transient behavior, despite the changed gains.

A straightforward and conspicuous strategy that an adversary may adopt is to completely nullify the integral gain ($k_i^{a}=0$), as illustrated in \cite{Cyb_attacks}. As shown in \cite{rsu}, the \ac{srf} \ac{pll} can still track the grid frequency with a proportional-only controller, that is, with the integral gain set to zero; however, exact phase tracking cannot be achieved, resulting in a steady-state error in power regulation when direct power control is implemented through current regulators, as in \cite{DSOGI}. Under this condition, an attacker can actively manipulate the active power output of the plant \cite{DSOGI}. Here, since $\eta^{s} = [(\omega_{g}-\omega_{n})/k_{p}^{a},\, 0] \neq [0,0]$, the attack is immediately detected: the induced phase error prevents the exact phase lock that synchronization requires, which is reflected in critical controller states such as the $dq$ voltages and triggers converter disconnection due to loss of synchronization, exposing the tampering. The more capable threat model of \cite{Exploiting_Inherent}, however, which considers all possible combinations of \ac{pll} gain manipulation that preserve integral action, always yields $\eta^{s}=0$, leaving no observable trace for operators to detect the attack, which makes it particularly dangerous and is therefore the threat model considered in this work. Since $\eta^{s}$ captures precisely any shift in the monitored steady-state values, it provides an exact mathematical characterization of the stealth notion introduced above.
\begin{remark}
    A \ac{pll} gain tampering cyber-attack remains stealthy if and only if $\eta^{s}=0$.
\end{remark}
\noindent Consequently, capturing any tampering of the \ac{pll} gains requires $\eta^{s} \neq 0$, a condition that, for the threat model of \eqref{eq:attack}, cannot be satisfied by the conventional \ac{pll} structure. This becomes clearer through \eqref{eq:srfequil}: owing to the exact phase tracking enabled by the integral term, the equilibrium point is always the same for a given grid frequency. This motivates a systematic analysis of the impact of both the frequency-estimation and synchronization \ac{pll} parameters on the small- and large-signal behavior of the \ac{ibr}, in order to anticipate gain-tampering strategies that degrade system performance and to design a detection mechanism capable of identifying such attacks.

%However, in extreme cases, such as the one considered in \cite{Cyb_attacks}, where the attacker completely nullifies the integral gain ($k_{i}^{a}=0$), then $\eta^{s}=[(\omega_{g}-\omega_{n})/k_{p}^{a}, \, 0 ]\neq [0,0]$. In this scenario, the integrator maintains its pre-attack value and the \ac{pll} still tracks the grid frequency. Importantly, $\eta^{s}$ is no longer zero, thereby enabling detecting the gain manipulation.
%Hence, there is a need to carefully reconstruct the \ac{pll} so that tampering of both the proportional and integral gains is reflected in the \ac{pll} equilibrium points.
%
\vspace{-2mm}
\section{PLL Impact on Stability and Transient Response}
\vspace{-1mm}
As evident from \eqref{eq:p_ref}, the frequency estimator and its interaction with nonlinear elements, such as \acp{rl}, critically influence the dynamic response and small-signal behavior of \ac{gfl} converters. To capture these effects, this section conducts an explicit small-signal stability analysis, complemented by a comprehensive simulation study that examines the relationship between the quality of grid-supporting services provided by the \ac{ibr} and the frequency estimator. 

\begin{figure}[b]
    \centering
    \includegraphics[width=0.9\linewidth]{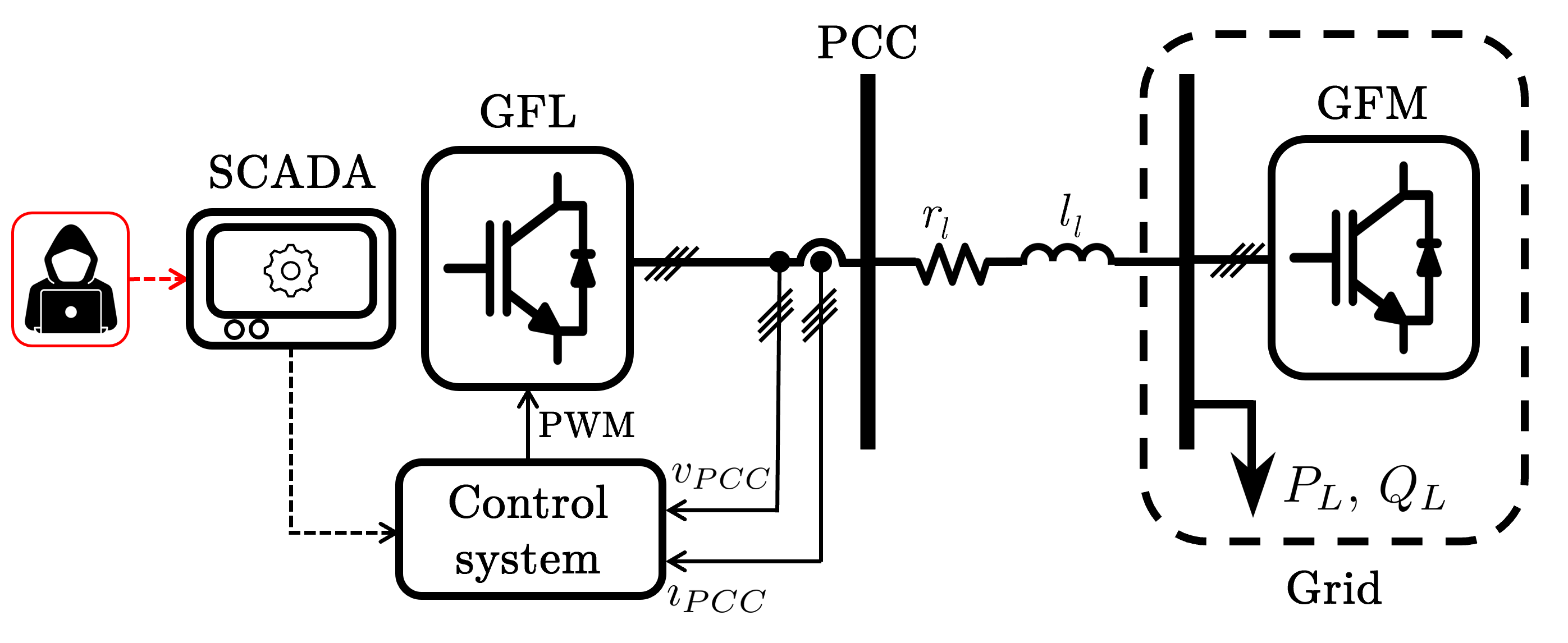}
    \vspace{-2mm}
    \caption{\ac{gfl} converter connected to a dynamic grid.}
    \label{fig:gfm_gfl}
       % \vspace{-6mm}
\end{figure}

\begin{figure*}[t!]
    \centering
    \begin{subfigure}{0.49\textwidth}
     \includegraphics[width=\linewidth]{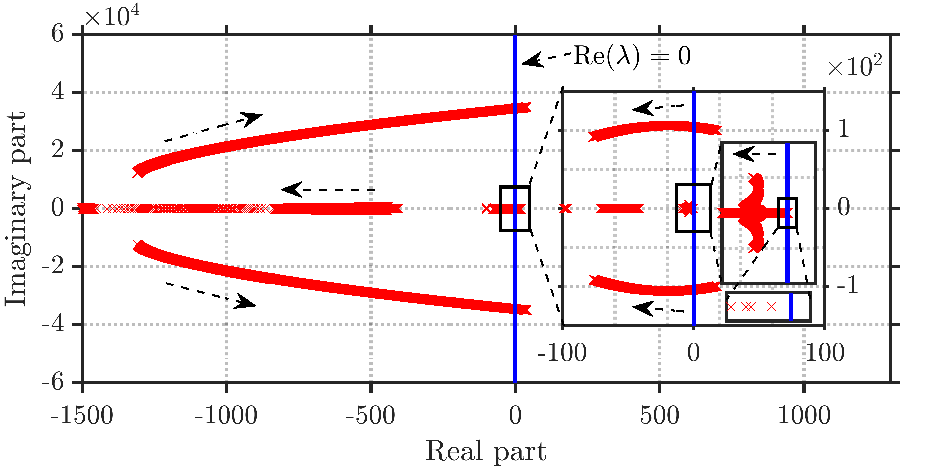}
    \caption{Synchronization \ac{pll} bandwidth: 40~Hz} 
    \label{fig:weak_fast}
    \vspace{-1mm}
    \end{subfigure}
    \begin{subfigure}{0.49\textwidth}
        \centering
     \includegraphics[width=1\linewidth]{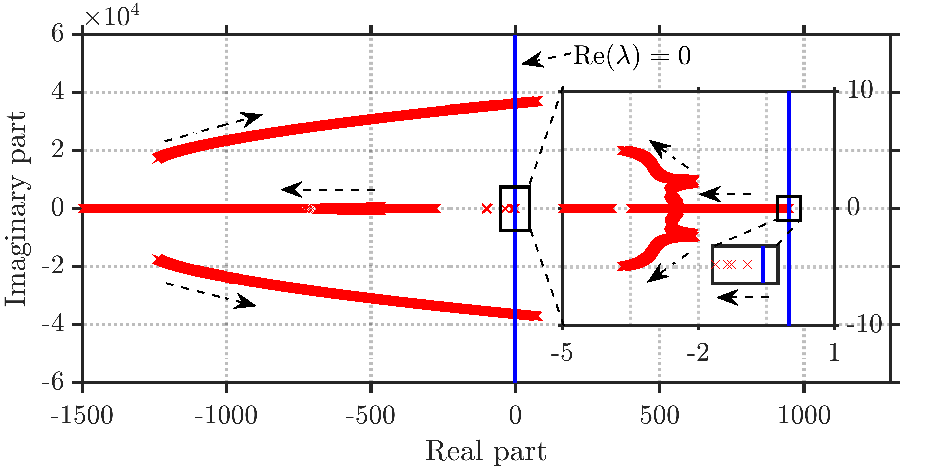}
    %\vfill
    \caption{Synchronization \ac{pll} bandwidth: 1~Hz} 
    \label{fig:weak_slow}
    \vspace{-1mm}
    \end{subfigure}
    \caption{Modal analysis for a \ac{gfl} converter connected to a \ac{gfm} converter through a 2.5~mH line, for different \ac{pll} bandwidths.} 
    \label{fig:weak}
    \vspace{-6mm}
\end{figure*}
\vspace{-2mm}
\subsection{Small-Signal Analysis}
\vspace{-1mm}
To analyze the interaction between the frequency-supporting \ac{gfl} converter and the grid,  a simplified single-bus model is considered, in which a \ac{gfm} converter is connected to both a \ac{gfl} converter and a load at the same bus. The simplified swing equation can be expressed as:
\begin{align}
    J\Delta \dot \omega = -D\Delta \omega + P_{GFM} + P_{GFL} - P_{L},
\end{align}
where $\Delta \omega = \omega - \omega_{n}$ is the frequency deviation of the \ac{gfm} converter, and $P_{GFM}$, $P_{GFL}$, and $P_{L}$ denote the power outputs of the \ac{gfm}, \ac{gfl} converters, and the load, respectively. The parameter $D$ is the droop coefficient of the \ac{gfm} \ac{ibr} and $J$ its virtual inertia.

Although modern grid codes require \ac{gfl} converters to follow the reference structure in~\eqref{eq:p_rl}, here~\eqref{eq:p_ref} is used for analytical clarity. Assuming that the \ac{pll} input voltage is $1~\mathrm{pu}$, the $q$-axis voltage used by the \ac{pll} outer loop in the local reference frame becomes:
\begin{align}
    v_{q}^{Lo} = \sin\!\Delta \phi.
\end{align}
where $\Delta \phi$ is given by:
\begin{align}
    \Delta \dot \phi = \omega - \omega_{o},
\end{align}
the \ac{pll} dynamics can be written as:
\begin{align}
    \Delta \dot \phi &= \Delta \omega - k_{p,o}\sin\Delta \phi - z_o,\\
    \dot z_o &= k_{i,o} \sin\Delta \phi.
\end{align}
where the superscript $(\cdot)^{n}$, denoting pre-attack quantities, is omitted for clarity.\\
Substituting these dynamics into the swing equation
%\begin{equation}
%\begin{aligned}
%\!\!J\!\Delta \dot\omega \!=\!\! -\!D\!\Delta\omega \!\!+\!\! P_{GFM}\! \!+\!\! P_s \!\!-\!\! P_L \!\!-\!\! m_P(k_{p,o} \!\sin \! \Delta\phi \!+\!\! k_{i,o} z_o\!),
%\end{aligned}
%\end{equation}
and linearizing the trigonometric terms around the equilibrium $\Delta \phi^{*}=0$, the closed-loop linearized system is:
\begin{align}
   \! \!\!J\!\Delta\dot{\tilde\omega} \!&=\!\! -\!D\Delta\tilde\omega \!+\!\! P_{GFM} \!+\!\! P_s \!\!-\!\! P_L \!\!-\! m_{\!P}(\!k_{p,o}\Delta \tilde \phi + k_{i,o} \tilde z_o\!),  \label{eq:sw_lin}\\
    \Delta \dot{\tilde\phi} &= \Delta \tilde \omega - k_{p,o}\Delta \tilde \phi - k_{i,o}\tilde z_o, &&\label{eq:df_lin}\\
    \dot {\tilde{z}}_o &= \Delta \tilde\phi. && \label{eq:z_lin}
\end{align}
The state matrix of this system is:
\[
\begin{bmatrix}
    -D/J & -m_{P}k_{p,o}/J & -m_{P}k_{i,o}/J \\
    1 & -k_{p,o} & -k_{i,o} \\
    0 & 1 & 0
\end{bmatrix}.
\]

Applying the Routh-Hurwitz criterion to obtain the condition for the eigenvalues to have strictly negative real parts yields the stability condition:
\begin{align}
    \!\!\frac{m_{\!P}k_{i,o}}{J} \!\!-\!\! k_{p,o}k_{i,o}
    \!\!< \!\!\frac{k_{p,o}^{2}D}{J} \!\!+\!\! \frac{m_{\!P}k_{p,o}^{2}}{J}
    \!\!+\!\! k_{p,o}\!\!\left(\!\!\frac{D}{J}\!\!\right)^{\!\!2}\!\!
    \!+\!\! \frac{Dm_{\!P}k_{p,o}}{J^{2}}. \label{eq:stab_init}
\end{align}

This inequality is not satisfied by every combination of \ac{pll} or network parameter values, meaning the system can become unstable for poorly chosen parameters. Notably, when $k_{p,o}=0$, a conventional \ac{pll} would be marginally stable, yet under this simplified model the overall system becomes unstable.
If instead the \ac{gfl} converter is connected to an infinite bus, $J\approx \infty$, the stability condition reduces to $-k_{p,o}k_{i,o} < 0,$
%\begin{align}
%    -k_{p,o}k_{i,o} < 0,
%\end{align}
which is always satisfied. In this case, the one-bus model explicitly reveals that the two subsystems decouple, guaranteeing system stability independently of their interaction.
To obtain a more interpretable stability condition, \eqref{eq:stab_init} is rewritten as:

\vspace{-2.5mm}
\begin{align}
    \!\!\!\frac{m_{P}k_{i,o}}{J}
    \!\!<\!\! k_{p,o}\!\!\left(\!\!k_{i,o}\!+ \!\frac{k_{p,o}D}{J} \!+ \!\frac{m_{P}k_{p,o}}{J}\!
    +\! \!\left(\frac{D}{J}\right)^{2}\!\!\!
    \!+\! \frac{Dm_{P}}{J^{2}}  \!\!\right)\!\!.
\end{align}
This inequality indicates that when the \ac{pll} is extremely under-damped, the combined system may readily become unstable. The simplified \ac{gfm} model is stable in stand-alone operation; therefore, any potential instability arises solely from the \ac{gfl} converter and its outer-loop \ac{pll}. A similar observation was reported in~\cite{pll_influence}, where interactions between \ac{gfm} and \ac{gfl} converters at low bandwidths were shown to induce instability. This instability is strongly influenced by the frequency estimator dynamics and may be exploited by adversaries to degrade system performance or, in extreme cases, destabilize the power system.

Although the previous results indicate strong coupling among the \acp{ibr}, the reduced-order model cannot fully capture their interactions, owing to simplifying assumptions such as neglecting essential \ac{ibr} dynamics, including the current and power control loops of the primary (voltage-frequency) control shown in Fig.~\ref{fig:gfl_control}, and, most importantly, the impact of the synchronization \ac{pll}. To address this limitation, an eigenvalue analysis is performed on the testbench shown in Fig.~\ref{fig:gfm_gfl} under both strong and weak grid conditions, using the complete nonlinear closed-loop model established in Sections~II and~III, incorporating the network dynamics and the current and power control loops. The eigenvalues are obtained directly from the Jacobian of this model, evaluated at the system's equilibrium point \cite{dq_pll}. The \ac{gfl} converter is rated at 0.1~MVA, whereas the \ac{gfm} \ac{ibr} has a rating of 1~MVA, with all droop coefficients for both converters set to 1\%.

\begin{figure}[t]
    \vspace{-3mm}
    \centering
    \includegraphics[width=.98\linewidth]{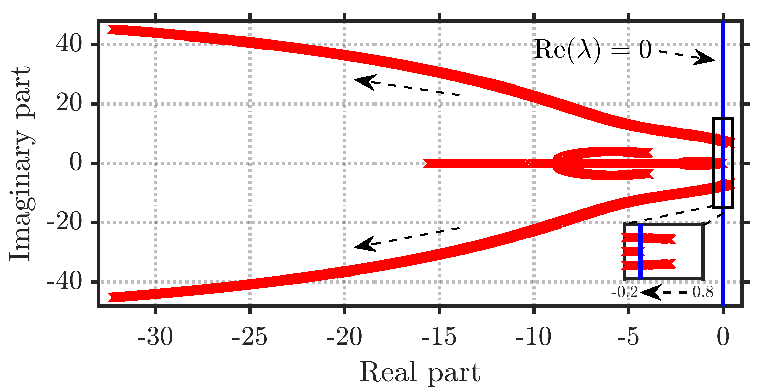}
    \vspace{-2mm}
    \caption{Dominant eigenvalues for a reduced controllers' bandwidth.}
    \label{fig:sl_cont}
    \vspace{-5mm}
\end{figure}

First, a weak-grid is considered with a 2.9 SCR index. The synchronization \ac{pll} bandwidth is set to 40~Hz, and the eigenvalues of the system are computed as the bandwidth of the frequency estimator is increased from 0 to 10~Hz. The damping ratio $\zeta$, is fixed at $1/\sqrt{2}$. The results are shown in Fig.~\ref{fig:weak_fast}, where the arrows indicate the migration of the eigenvalues as the \ac{pll} bandwidth increases. It can be observed that the system is unstable for lower bandwidth values, primarily due to weak-grid conditions associated with the low SCR.
As the estimator bandwidth increases, the frequency-support controller stabilizes the system. This stabilizing effect is attributed to the grid-forming characteristics of the controller and its contribution to regulating the \ac{pcc} frequency and voltage. However, further increasing the bandwidth eventually destabilizes the system, as the power and current controllers as well as the \ac{pll} are unable to follow in time the output of the ancillary services controller. To further illustrate this phenomenon, the controller bandwidths are reduced by a factor of three, and the same analysis is repeated. The results, shown in Fig.~\ref{fig:sl_cont}, indicate that without the frequency-support mechanism the system remains unstable; however, as the frequency estimator bandwidth increases, stability is achieved at a lower frequency compared to Fig.~\ref{fig:weak_fast}. Moreover, the system remains stable over a wider range of estimator bandwidths. Overall, excessively slow tuning of the frequency estimator limits the frequency-support capability of the \acp{ibr} and can lead to instability under weak-grid conditions, while overly aggressive tuning may produce similar destabilizing effects.
%
%This behavior can be explained by the time-separation principle underlying \ac{gfl} control design. This principle requires the outer control loops to operate at slower time scales than the inner current loops. Violation of this separation renders the system more susceptible to instability. Consequently, reducing the controller bandwidth mitigates high-frequency instabilities and improves the low-frequency small-signal response of the system.
%%
%
%
%

\begin{figure}[h]
    \vspace{-3mm}    
    \centering
    \includegraphics[width=.98\linewidth]{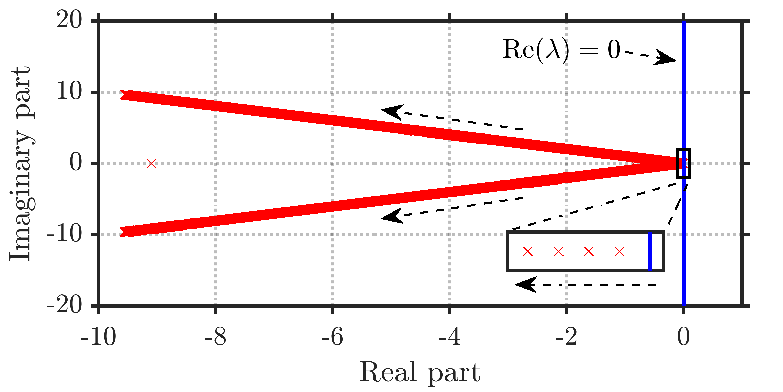}
    \vspace{-2mm}
    \caption{Effect of an infinite bus connection on dominant eigenvalues.}
    \label{fig:both}
    \vspace{-5mm}    
\end{figure}

Next, the synchronization \ac{pll} bandwidth is reduced to 1~Hz, and the experiment is repeated. The results in Fig.~\ref{fig:weak_slow} demonstrate that this adjustment mitigates both small-signal instability and low-frequency oscillatory modes. In contrast to using the synchronization \ac{pll} for frequency estimation, increasing the bandwidth of a dedicated \ac{pll} now enhances the fault-ride-through capability of the \ac{ibr}, as the dominant eigenvalues shift further into the left-half plane. However, further increasing the \ac{pll} bandwidth again leads to instability. Finally, the transmission line is removed and the \ac{gfm} converter is replaced by an infinite bus. The analysis is repeated for both \ac{pll} bandwidth configurations, and the dominant eigenvalues are shown in Fig.~\ref{fig:both}. As expected, the dominant eigenvalues coincide in both cases, and the system remains stable for all \ac{pll} bandwidths.

These results underscore the critical role of frequency support and estimator dynamics in system stability. Improper gain tuning can degrade performance or even induce instability, providing a potential avenue for adversaries to stealthily compromise the small-signal behavior of \acp{ibr} and the broader power system.

\begin{comment}
Equally important to studying its stability, is analyzing the interaction of the controller with the frequency estimator. To this end, we replace the frequency estimation \ac{pll} equations to the controller and ignore the rate-limiter and dead-zone for simplicity:
\begin{align}
    P_{r}=P_{s}-m_{P}\left(k_{p,o}v_{c,q}^{Lo}+z_{o}\right), \label{eq:actp_rep}
\end{align}
An approximation of the power-output of the converter can be given through:
\begin{align}
    P=\frac{V_{i}V_{g}}{X}\sin(\theta_p+\theta_{c}-\theta_{pcc})
\end{align}
where $\theta_{c}$ and $\theta_{pcc}$ correspond to the phase of the inverter and the \ac{pcc} voltage. Assuming that the angle inside the sine wave is small, this formulation can be approximated by
\begin{align}
    P\approx\frac{V_iV_g}{X}(\theta_p+\theta_{c}-\theta_{pcc})
\end{align}
Solving for $\theta_{pcc}$ and assuming $V_i\approx V_g=1$~pu:
\begin{align}
    \theta_{pcc}\approx\frac{P}{V_iV_g}-\theta_p-\theta_{c}
\end{align}
Replacing the active power output with its reference, as given by \eqref{eq:actp_rep} results in:
\begin{align}
    \theta_{pcc}\approx X\left(P_{s}-m_{P}\left(k_{p,o}v_{c,q}^{Lo}+z_{o}\right)\right) -\theta_p-\theta_{c}
\end{align}
\end{comment}

\subsection{Large-Signal Analysis}
\begin{figure}[h!]
    \centering
    \includegraphics[width=0.98\linewidth]{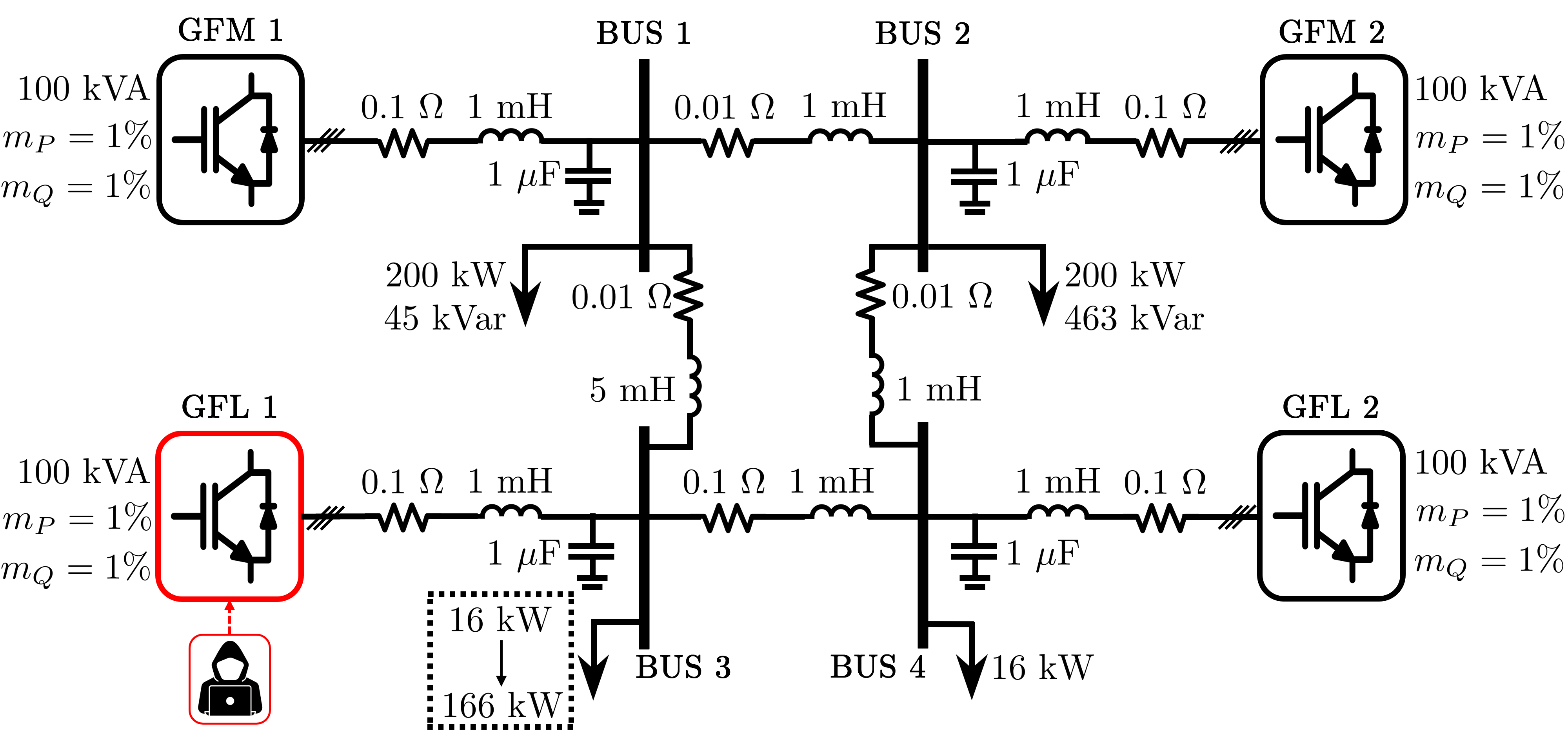}
    \caption{Simulation benchmark.}
    \label{fig:placeholder}
        %\vspace{-6mm}
\end{figure}

%\vspace{-1mm}
This section investigates the impact of the frequency estimator dynamics on the performance of the grid-supporting \ac{ibr} during large-signal disturbances, such as load changes. %These are the main scenarios where grid-codes require grid-connected resources to provide active and reactive power support to assist operators in maintaining frequency and voltage states close to their nominal values. 
To conduct a more holistic analysis, a grid with four \acp{ibr} is considered, as shown in Fig. \ref{fig:placeholder}.

\begin{figure*}[h!]
    \centering
    \begin{subfigure}{0.49\textwidth}
     \includegraphics[width=\linewidth]{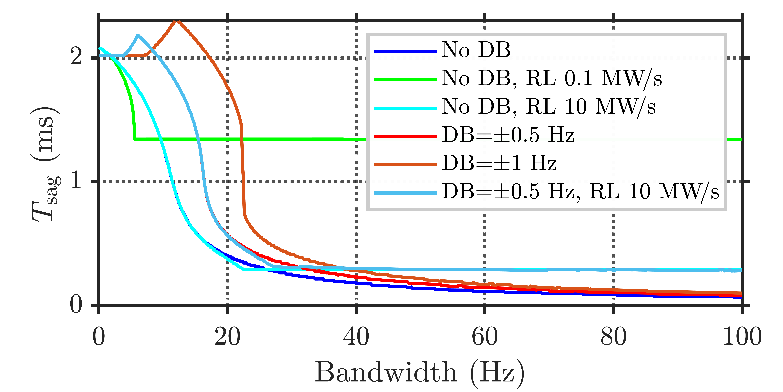}
    \vspace{-5mm}
    \caption{Inductive network.} 
    \label{fig:v_graph}
    \end{subfigure}
    \begin{subfigure}{0.49\textwidth}
        \centering
     \includegraphics[width=1\linewidth]{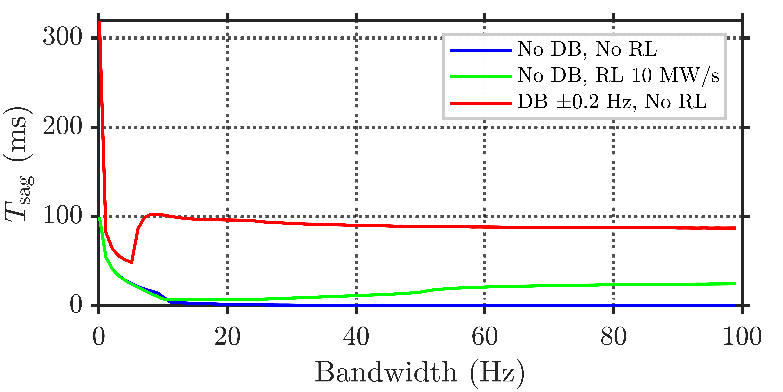}
    \vspace{-5mm}
    \caption{Resistive network.} 
    \label{fig:vres_graph}
    \end{subfigure}
    \vspace{-2mm}
    \caption{Voltage sag duration for different controller and \ac{pll} parameters.} 
    \label{fig:vdn_graph}
    \vspace{-6mm}
\end{figure*}

%\begin{figure}[t]
%    \centering
%    \includegraphics[width=\linewidth]{figs/vdrop.eps}
%    \caption{Voltage sag duration for different controller parameters in an inductive network.}
%    \label{fig:v_graph}
%\end{figure}

The performance of the frequency-support mechanism is evaluated by measuring the duration of a voltage sag induced by a significant load change at bus~3 ($+0.15$~MW), where the under-study grid-supporting \ac{ibr} is located. % All system parameters are kept constant, except for the configuration of the primary controller of GFL~1 and the bandwidth of its frequency estimator.
Specifically, the voltage-sag duration is measured for different deadband settings, \ac{rl} derivative limits, and frequency-estimator bandwidths. The results are illustrated in Fig.~\ref{fig:v_graph}.

Each primary controller configuration yields a distinct dynamic response, which may either improve or degrade the voltage recovery time. The shortest recovery times are obtained when no deadband is applied, allowing the controller to respond immediately to the disturbance. Introducing a \ac{rl} with tight power-derivative constraints significantly improves the response for low \ac{pll} bandwidths by suppressing excessively aggressive control actions. For higher bandwidths, the recovery time remains nearly constant, as the high-derivative response of the primary controller is constrained by the \ac{rl}. Once the \ac{rl} becomes inactive, the frequency estimator has already converged to the grid frequency and therefore has no further impact on the short-term system dynamics. Increasing the derivative limits results in similar behavior; however, the saturation of the recovery time occurs at a substantially shorter duration.

Incorporating a deadband in the controller limits preemptive control actions and reduces the participation of the \ac{ibr} in grid-support services. As observed in Fig.~\ref{fig:v_graph}, the converter does not contribute to voltage recovery when the estimator bandwidth is below 4~Hz and the deadband is set to $\pm0.5$~Hz. For bandwidths between 4 and 9~Hz, the controller operates counter-productively and increases the recovery time, whereas for higher bandwidths it improves the response. Increasing the deadband to $\pm1$~Hz leads to qualitatively similar but more pronounced effects. The combined use of \acp{rl} and deadbands produces an interaction in which the controller degrades performance at low estimator bandwidths, improves recovery at intermediate bandwidths, and ultimately leads to saturation of the recovery time at high bandwidths due to the restrictive action of the \ac{rl}.

The impact of the frequency estimator is more pronounced in resistive networks. To further examine this effect, the experiment is repeated for a predominantly resistive system ($X/R=0.6$), where both lines and loads exhibit mainly resistive characteristics. As shown in Fig.~\ref{fig:vres_graph}, the frequency estimator plays a critical role in disturbance rejection, with recovery times ranging from 0.1~ms to 300~ms. However, its interaction with nonlinear protection elements may either improve or degrade recovery performance. This behavior is most evident when the grid-support controller employs \acp{rl}: lower bandwidths reduce recovery time, whereas, in contrast to inductive networks, increasing the bandwidth results in a gradual increase in recovery time.
%\begin{figure}[h!]
%    \centering
%    \includegraphics[width=1\linewidth]{figs/vdrop_dn.eps}
%    \caption{Voltage sag duration under different ancillary services controller, in a resistive network.}
%    \label{fig:vdn_graph}
%\end{figure}
These results highlight the influence of the frequency estimator and its dynamic response on fast frequency support services. While the combination of conventional primary control with nonlinear and discontinuous elements can improve system performance during disturbances, improper tuning may instead degrade the response. More importantly, deliberate or \textit{malicious} manipulation of controller parameters can render the frequency-support mechanism counterproductive and may lead to instability of the \ac{ibr} as well as adverse interactions with surrounding loads and resources.

%Nevertheless, the small-signal analysis indicates that even under weak-grid conditions, properly tuned grid-supporting \ac{gfl} \acp{ibr} remain well within stable operating margins; therefore, a successful cyber-attack must \textbf{significantly} alter the proportional and integral gains of the \ac{pll} to compromise system stability.
%
%
\vspace{-2mm}
\section{\ac{rsu}-Based Detection of PLL Gain Tampering}
\begin{figure}[t]
    \centering
    \vspace{-1mm}
    \includegraphics[width=0.98\linewidth]{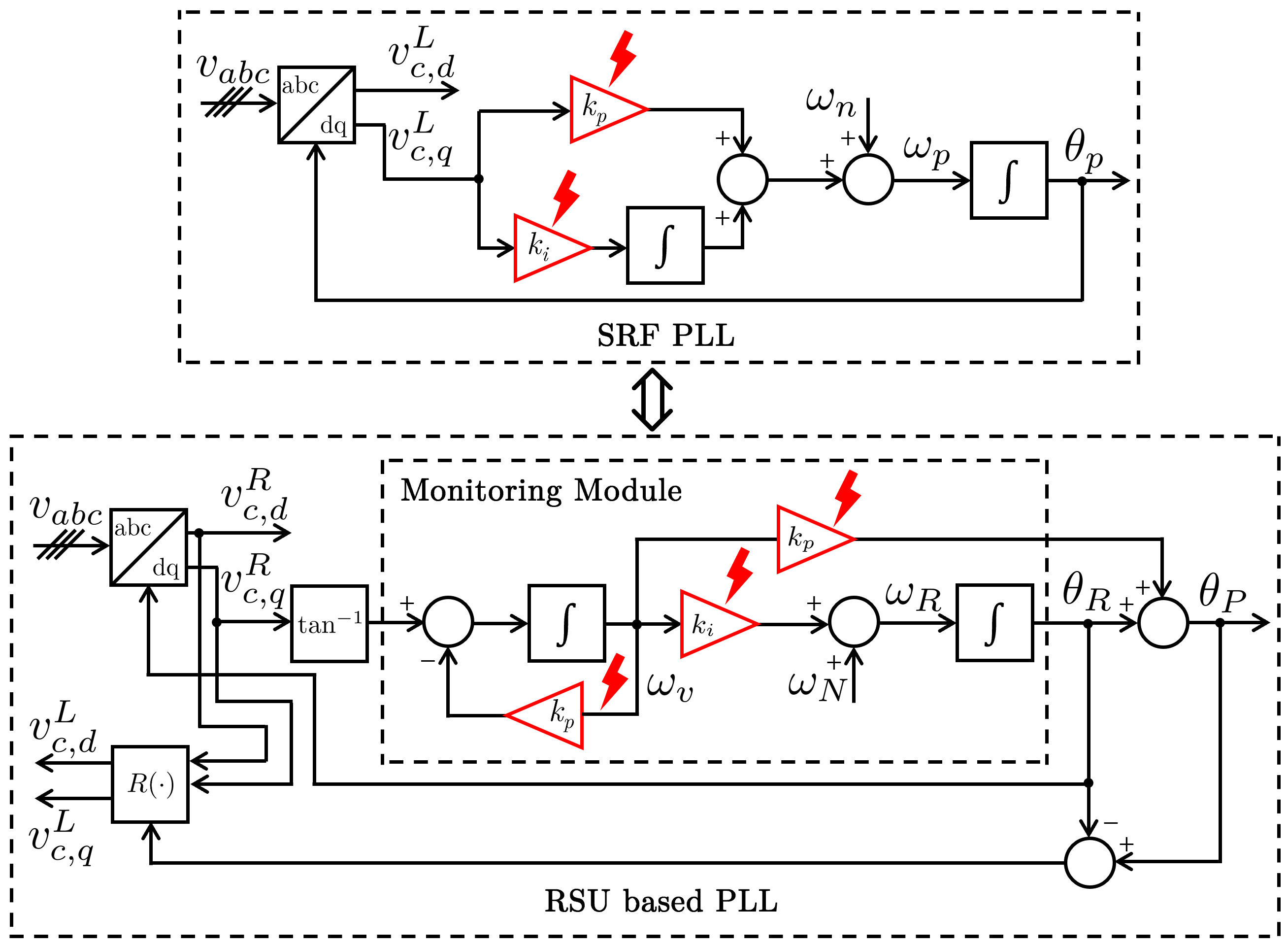}
    \vspace{-2mm}
    \caption{Proposed \ac{rsu}-based \ac{pll}.}
    \label{fig:RSUPLL}
        \vspace{-6mm}
\end{figure}

%\vspace{-1mm}
A key conclusion derived from the threat-model impact assessment is that, to meaningfully affect the small-signal behavior of the plant, an adversary must significantly alter the proportional gain, the integral gain, or both, since properly tuned \acp{pll} yield a system that is neither marginally stable nor underdamped. Building on this insight, this section presents an effective detection mechanism based on the \ac{rsu} framework introduced in \cite{rsu}, capable of capturing stealthy \ac{pll} gain manipulations without compromising system performance. The proposed detection scheme relies on observing the internal states of an alternative \ac{pll} implementation that behaves identically to the conventional \ac{srf} \ac{pll}. The proposed \ac{pll} is described by:

%
%
%A key conclusion acquired by the threat-model impact assessment is that significantly altering either the proportional or integral gains is required to meaningfully impact the small-signal behavior of the plant, since properly tuned \acp{pll} yield a system that is neither marginally stable nor underdamped. Building on this insight, this section presents an effective detection mechanism based on the \ac{rsu} framework introduced in \cite{rsu}, capable of capturing stealthy \ac{pll} gain manipulations without compromising system performance. The detection relies on observing the internal states of the alternative \ac{pll} implementation, which behaves identically to the conventional \ac{srf} \ac{pll}. The proposed \ac{pll} is described by
%
\begin{align}
    \dot \theta_{R} &=\omega_R= \omega_{N} + k_{i} z, \label{eq:rsu1} \\
    \dot z &= \arctan v_{c,q}^{R} - k_{p} z, \label{eq:rsu2}
\end{align}
with the Park transformation phase angle for the local control frame given by
\begin{align}
    \theta_{P} = \theta_{R} + k_{p} z.
\end{align}
Here \(\omega_{N}\) is a tunable constant, and the superscript \((\cdot)^{R}\) denotes quantities expressed in the \ac{rsu} internal reference frame. Additionally, $\omega_{R}$ and $\theta_{R}$ are \ac{rsu} related internal states. An explicit illustration of the proposed module is presented in Fig. \ref{fig:RSUPLL}. Considering the coordinate transformation in \eqref{eq:tran}, the pre-attack equilibrium point of the \ac{rsu} is:
\begin{align}
    \Delta \theta^{\rm s} = k_{p}^{n} (\omega_{g} - \omega_{N}) / k_{i}^{n}, \quad z^{\rm s} = (\omega_{g} - \omega_{N}) / k_i^{\rm n}, \label{eq:equila}
\end{align}
whereas now $\Delta \dot\theta=\omega_{g}-\omega_{R}$. Note, at steady state, the phase error between the Park transformation phase and the grid voltage, $\int (\dot \theta_{P} - \omega_{g}) dt$, is zero. In contrast, due to the \ac{rsu} structure, the internal phase error, $\arctan v_{c,q}^{R}=\Delta \theta$, is nonzero. Note that $\omega_{N}$ is a constant feed-forward term, traditionally set to $\omega_n$, and therefore does not affect the small-signal dynamic response of the \ac{pll}. Conventionally, it reduces the tracking effort of the \ac{pll} by providing a frequency reference close to that of the grid; here, however, $\omega_N$ assumes two additional roles that reinforce the proposed detection framework against the adopted threat model, as detailed below.

Evidently, the vector $\eta^{s}$ now captures any \ac{pll} gain tampering, since the equilibrium points of \eqref{eq:rsu1} and \eqref{eq:rsu2} depend on both $k_{p}^{n}$ and $k_{i}^{n}$, leaving no exploitable pattern for an adversary to mask an attack. Furthermore, any change in the controller gains induces a transient in the \ac{ibr} response, so that even under communication tampering an anomalous, unexplained event is flagged. Beyond this, $\omega_N$ serves two roles. First, it renders the communicated signals privacy-preserving: with $\omega_{N} \neq \omega_n$, \eqref{eq:equila} provides only two equations in the three unknowns $k_{p}^{n}$, $k_{i}^{n}$, and $\omega_{N}$, so the gains cannot be fully reconstructed by an interceptor without knowledge of $\omega_{N}$; the plant operator, which alone knows the implementation structure and $\omega_{N}$, remains the only entity able to recover them. Second, it sets the detection margin: since $\omega_N$ scales the equilibrium points in \eqref{eq:equila}, increasing $|\omega_g - \omega_N|$ enlarges the steady-state offsets $|\Delta\theta^{\rm s}|$ and $|z^{\rm s}|$, and hence the equilibrium shift produced by a gain change, making gain tampering easier to distinguish from frequency fluctuations, measurement noise, and grid-side disturbances. Since $\omega_N$ does not affect the transient response of the \ac{pll}, it can be tuned to achieve the desired detection margin and is set once at startup. Moreover, it need not be exposed to the supervisory (SCADA) interface, which keeps it private and beyond the reach of the adversary, who could otherwise mask only integral-gain tampering; since $\Delta\theta^{\rm s}=k_p^n z^{\rm s}$ from \eqref{eq:equila} fixes the proportional relation between the two equilibria independently of $\omega_N$, any change in $k_p^n$ alters this relation and exposes the attack. Alternatively, $\omega_N$ can be periodically re-keyed through a cryptographic scheme, so that a leaked value can be exploited, for gain inference via \eqref{eq:equila} or attack masking, only until its next update.

The proposed scheme thus produces an equilibrium-based footprint that exposes any change in the \ac{pll} gains, prompting and enabling operators to investigate whether the change reflects malicious tampering or a legitimate manual or adaptive reconfiguration. This distinction is ultimately made by the plant operator; the proposed device reveals what was previously invisible, owing to the stealthy nature of the attack established in Section~IV. Detection is governed by two parameters: an observation window $T_w$, which sets the interval over which a change in the equilibrium is sought, and a dwell time $T_d$, which sets how long that change must persist before it is declared significant. Both depend on the response time of the conditioning filters and the settling time of the slowest \ac{pll} the plant operator may employ, and the dwell time should exceed the duration of the transients expected under both normal and adverse grid conditions, while the magnitude and clarity of the equilibrium offset can be optimized by tuning $\omega_N$. Accordingly, there is a trade-off: a small dwell time yields early detection but may register transients as tampering, whereas a larger dwell time delays detection but ensures no false positives occur. However, since the impact of this attack manifests only during significant transients, which are unlikely to occur within even a few seconds of the attack, a more conservative dwell time is generally preferable. Moreover, since the equilibrium points are functions only of the \ac{pll} gains and the grid frequency, the latter being practically constant over the timescale of the attack, a sustained equilibrium offset can arise solely from a gain change, thereby precluding false positives.

The algorithmic equivalent of the detection pipeline is presented in Algorithm~\ref{alg:rsu_detection}. At every sample, the equilibrium vector $\eta^{\rm s}$ is compared with its value one observation window $T_w$ earlier, and whenever a component of the resulting difference persistently exceeds its corresponding threshold, $\epsilon_{\Delta\theta}$ or $\epsilon_{z}$, for longer than the dwell time $T_d$, the associated flag, $\sigma_{\Delta\theta}$ or $\sigma_{z}$, is set to one, signifying a gain-change event. The thresholds and timing parameters must be selected relative to the noise level, the filtering parameters, and, most importantly, the steady-state levels of the equilibrium points. To accommodate possible gain adaptation, the observation window can be referenced to the slowest \ac{pll} the operator or adaptation algorithm may employ and conservatively set to a few seconds. Even such a conservative window detects the tampering well before it can escalate into harmful operation, whereas a less conservative design, tuned to the neighborhood of the actual \ac{pll} bandwidth in use, reduces the detection delay to a few hundred milliseconds. A malicious gain tampering may go unnoticed only when it produces an equilibrium change smaller than $\epsilon_{\Delta\theta}$ or $\epsilon_{z}$, which occurs for marginal gain changes that in turn have a negligible impact on the \ac{pll} response, thereby ensuring that any tampering capable of degrading system performance is reliably detected.

\begin{algorithm}[t]
\caption{RSU-Based Detection of PLL Gain Tampering}
\label{alg:rsu_detection}
\begin{algorithmic}[1]

\State \textbf{Constants:} $\omega_N$, sampling time $T_s$
\State \textbf{Input:} equilibrium vector $\eta^{\rm s} = (\Delta\theta^{\rm s},\, z^{\rm s})$; window $T_w$; dwell time $T_d$; thresholds $\boldsymbol{\epsilon} = (\epsilon_{\Delta\theta},\, \epsilon_{z})$
\State \textbf{Output:} flags $\boldsymbol{\sigma} = (\sigma_{\Delta\theta},\, \sigma_{z})$
\State $\boldsymbol{f} \gets \boldsymbol{0}$, \quad $\boldsymbol{\sigma} \gets \boldsymbol{0}$, \quad $w \gets T_w / T_s$
\State Filter $\eta^{\rm s}$ through the conditioning low-pass filter
\For{each sample $n$}
    \State $\Delta\boldsymbol{\eta} \gets \lvert \eta^{\rm s}(n) - \eta^{\rm s}(n-w) \rvert$ \Comment{element-wise windowed difference}
    \For{each component $i \in \{\Delta\theta,\, z\}$}
        \If{$\Delta\eta_i > \epsilon_i$}
            \State $f_i \gets f_i + T_s$ \Comment{accumulate dwell}
        \Else
            \State $f_i \gets 0$ \Comment{reset on transient}
        \EndIf
        \If{$f_i \geq T_d$}
            \State $\sigma_i \gets 1$ \Comment{gain-change event on component $i$}
        \EndIf
    \EndFor
\EndFor
\end{algorithmic}
\end{algorithm}

To demonstrate how the proposed design does not disturb the transient performance of the conventional \ac{srf} \ac{pll}, the transfer function between $\theta_{P}$ and $\theta_{g}=\int\omega_{g}dt$ is formulated:
\begin{align}
    \frac{\theta_{P}}{\theta_{g}}=\frac{k_{p}s+k_{i}}{s^2+k_{p}s+k_{i}},
\end{align}
which obviously coincides with that of the \ac{srf} \ac{pll}, contrary to the alternative synchronization designs presented in \cite{DSOGI, Cyb_attacks} that fundamentally alter the dynamics of the \ac{pll}. Therefore, the proposed and conventional \ac{srf} \acp{pll} are small-signal equivalent and exhibit identical small-signal responses. Consequently, since they share a common input-output relation, any small-signal stability analysis carried out for the \ac{srf} \ac{pll} yields exactly the same results for the proposed surrogate \ac{rsu}-based \ac{pll}, including that of Section~V. Moreover, given its low computational complexity, comprising the same number of states as the \ac{srf} \ac{pll}, and owing to this equivalence, the proposed structure can be readily integrated into existing \ac{gfl} configurations, augmenting them with the ability to detect malicious gain tampering while fully preserving the tracking performance of the original \ac{pll}.

Another important advantage of the proposed scheme over \cite{Cyb_attacks} is that it can be readily adopted to any \ac{gfl} control structure. Specifically, the controller in Fig.~\ref{fig:gfl_control} requires accurate current decoupling to actively isolate the active and reactive power responses. This decoupling relies on an explicit formulation of the Park transformation rotating frequency, which is not available in \cite{Cyb_attacks,DSOGI}. In contrast, for the proposed approach, the rotating frequency is given by:
\begin{equation}
\begin{aligned}
    \omega_{P} &= \omega_{N} + k_{i} z + k_{p} \dot z, \\
    &=\omega_{N} + k_{i} z +k_{p} \left(\arctan v_{c,q}^{R} - k_p z\right).\label{eq:wp}
\end{aligned}
\end{equation}

The proposed surrogate \ac{pll} can also be combined with the same \ac{qsg}-based positive- and negative-sequence separation employed by a conventional \ac{srf} \ac{pll} \cite{sogi}, thereby enabling operation under unbalanced networks and dominant low-order harmonics, such as the $5^{\rm th}$ and $7^{\rm th}$. The integration is a drop-in modification: the \ac{qsg} sequence-separation stage is retained unchanged, and only the synchronization-loop equations, i.e. \eqref{eq:theta}-\eqref{eq:zeta} are replaced by those of the \ac{rsu}-based \ac{pll} in \eqref{eq:rsu1}-\eqref{eq:rsu2}, with the frequency-adaptive resonators driven by the \ac{rsu} output frequency \eqref{eq:wp}. The small-signal equivalence between the two \acp{pll} is preserved under this implementation, since their small-signal input-output relation is identical and the \acp{qsg} derive the filtered positive- and negative-sequence signals from the output frequency of each \ac{pll}, which, as proven above, are small-signal equivalent. The same conditioned sequence signals are therefore presented to both loops, leaving the equivalence intact. 
\vspace{-3mm}
\subsection{Extension of the Detection Scheme to Multi-Layer IBR Control}
\vspace{-1mm}
Extending the proposed detection scheme to other \ac{ibr} controllers is straightforward and relies on the uniqueness of the equilibrium points of the power system. Unlike \acp{pll}, these controllers can be suitably modified such that any change in a controller gain produces a unique effect on the equilibrium point of the controllers’ internal states. Without loss of generality, the following alternative PI controller is considered:
%\vspace{-5mm}
\begin{align}
u = k_{p}\left(a x_{r} - x\right) + k_{i}\int \left(x_{r} - x\right),
\end{align}
%\vspace{-0.5em}
where $x$ and $x_{r}$ denote the measured and reference states, respectively, and $a \in (0,1)$ is a constant introduced to mask the controller’s equilibrium point.

Implementing this PI controller in place of a conventional one yields unique, nonlinear equilibrium-point mappings of the form:   
\begin{align}
\bfg 0 = \underbrace{\bfg g\left(k_{p}, k_{i}, x_{r}\right)}_{\text{Nonlinear function}},
\end{align}
which enable the detection of controller gain tampering by recovering the controller gains from the solutions of $\bfg g = \bfg 0$. Such gain recovery is feasible only for the plant operator, who, analogous to the case of the \ac{pll} and the parameter $\omega_{N}$, is the sole entity with knowledge of the controller structure and the parameters $\bfg a$. This property significantly complicates attempts by an adversary to mask equilibrium-point alterations, thereby enhancing the resilience of the \ac{srf} \ac{pll}-based control architecture against attacks targeting critical \ac{ibr} infrastructure.
\begin{comment}
\subsection{RSU-based Frequency Estimator}
Although using a \ac{pll} as a frequency estimator enables providing real-time frequency support to the grid, it can also induce oscillations, resonance and even instability. Furthermore, it offers redundant functions such as exact phase tracking that can restrain stability margins. Practically, a frequency estimator is required to strictly track the frequency of an input signal, smoothly and accurately without significant transients. This properties are conveniently offered by the \ac{rsu} and more importantly it offers significant advantages related to transient response over the \ac{srf} \ac{pll}. To better comprehend this, we consider the transfer function of the RSU:
\begin{align}
    \frac{\theta_{R}}{\theta_{g}}=\frac{k_{i}^{i}}{s^2+k_{p}^{i}s+k_{i}^{i}}.
\end{align}
Obviously, the RSU shares the same poles as the \ac{srf} \ac{pll} but it does not consist of any low-frequency zeros that tend to cancel out the response of dominant poles and induce oscillations. Furthermore, similarly to the swing equation, the RSU acts as a low-pass filter that dampens oscillations, and therefore, for the same settling time as the \ac{srf} \ac{pll}, it offers considerably better filtering. This allows to further decrease the settling time of the RSU compared to the \ac{srf} \ac{pll} and still display the same harmonic content as the latter synchronization device. 
\end{comment}

\vspace{-2mm}
\section{Experimental Results}
\begin{figure}[t]
    \centering
    \includegraphics[width=0.95\linewidth]{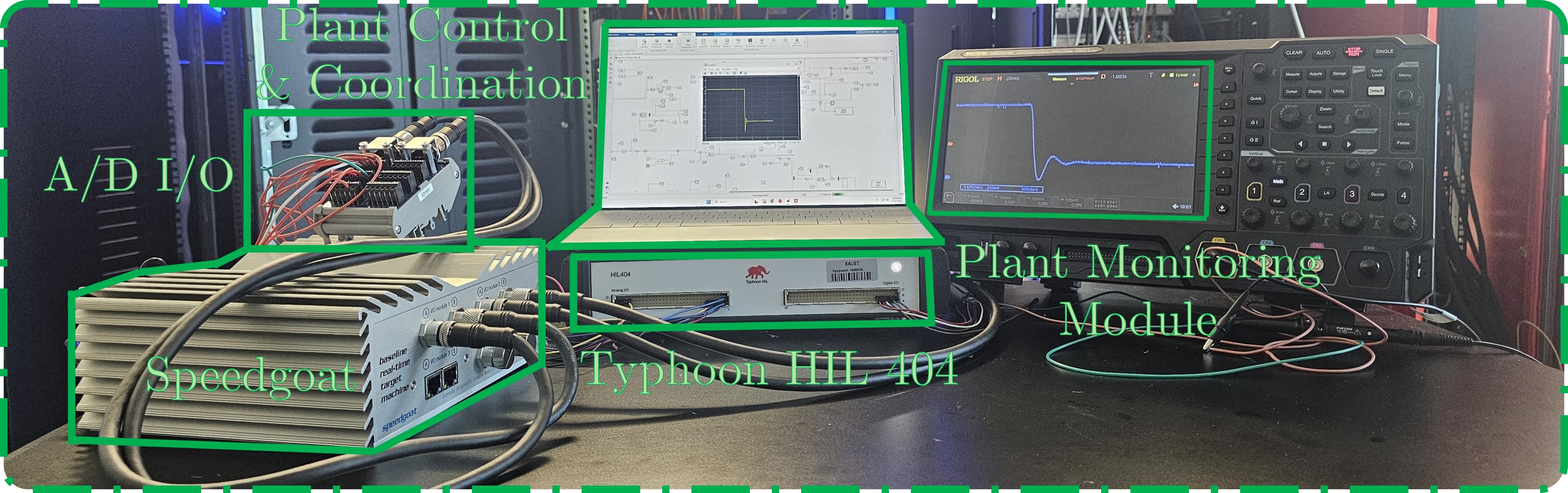}
    \caption{Experimental benchmark.}
    \label{fig:pic_exp}
        \vspace{-4mm}
\end{figure}

\vspace{-1mm}
This section presents an experimental validation of the proposed detection scheme under realistic operating conditions using a \ac{chil} setup, depicted in Fig.~\ref{fig:pic_exp}. Two scenarios are examined, under balanced and under unbalanced, distorted grid conditions, respectively. Each scenario establishes the small-signal equivalence between the \ac{rsu}-based and conventional \ac{srf} \acp{pll}, and further demonstrates the real-time applicability of the proposed \ac{pll} and its ability to detect the adopted threat model. The attack is applied to a \ac{gfl} \ac{ibr}, whose firing pulses are generated by a controller implemented on a Speedgoat Baseline real-time target, while the cyber-physical system is emulated in the Typhoon HIL 404 environment.

\vspace{-3mm}
\subsection{Operation and Attack Detection under Balanced Grid Conditions}
\vspace{-1mm}
The system shown in Fig.~\ref{fig:gfm_gfl} is considered, where a \ac{gfl} converter is connected to a grid emulated by a \ac{gfm} converter. The critical system and controller parameters are summarized in Table~\ref{tab:gfl_gfm_params}. The synchronization \ac{pll} bandwidth is set to 20~Hz and the frequency-estimation bandwidth to 2.5~Hz, with $\zeta = 1/\sqrt2$ and $f_{N} = 49.9~\mathrm{Hz} = \omega_{N}/2\pi$. Furthermore, the detection algorithm parameters are set to $\epsilon_{\Delta\theta} = 2\times10^{-2}$, $\epsilon_{z} = 5\times10^{-4}$, and $T_d = T_w = 0.8$~s. The two objectives are to demonstrate that the proposed \ac{pll} implementation behaves identically to the conventional \ac{srf} \ac{pll}, and to verify the effectiveness of the detection framework. All presented signals are filtered using a low-pass filter with a 100~Hz cut-off frequency.
\begin{table}[t]
%\vspace{-mm}
\caption{Experimental testbench parameters.}
\vspace{-2mm}
\label{tab:gfl_gfm_params}
\centering
\begin{tabular}{lcc}
\hline
\textbf{Parameter} & \textbf{GFL Inverter} & \textbf{GFM Inverter} \\
\hline
Power Rating & 0.1~MVA & 1~MVA \\
Frequency Droop & 5\% & 2\% \\
Voltage Droop & 5\% & 5\% \\
\hline
\multicolumn{3}{c}{\textbf{Common System Parameters}} \\
\hline
Base Load &   \multicolumn{2}{c}{\hfill 0.3~MVA}\\
Line &  \multicolumn{2}{c}{\hfill 0.01 $\Omega$, 0.2~mH}\\
Switching Frequency & \multicolumn{2}{c}{\hfill 15~kHz} \\
Nominal Line Voltage & \multicolumn{2}{c}{\hfill 400~V} \\
\hline
\end{tabular}
\vspace{-2mm}
\end{table}

\begin{figure}[t]
\vspace{-2mm}
    \centering
    \begin{subfigure}{0.45\textwidth}
     \includegraphics[width=\linewidth]{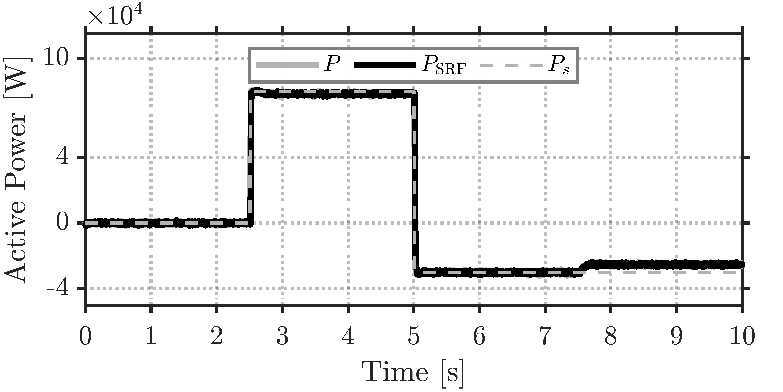}
     \captionsetup{margin={2mm,0mm}}
    \caption{} 
    \label{fig:oper_exp_pw}
    \end{subfigure}
    \begin{subfigure}{0.45\textwidth}
    \centering
     \includegraphics[width=1\linewidth]{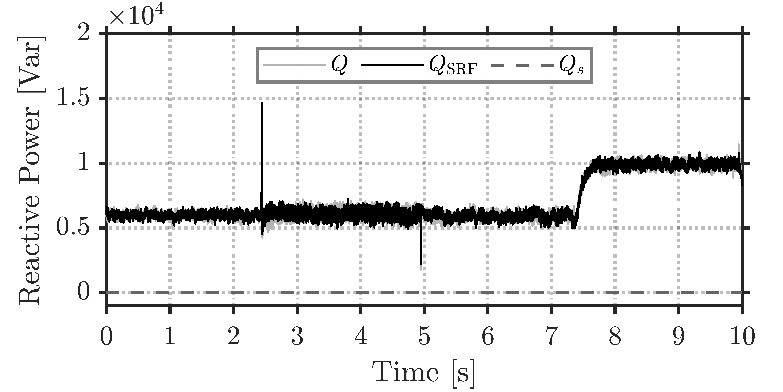}
     \captionsetup{margin={2mm,0mm}}
    \caption{} 
    \label{fig:oper_exp_rpw}
    \end{subfigure}
    \begin{subfigure}{0.45\textwidth}
     \centering
     \includegraphics[width=1\linewidth]{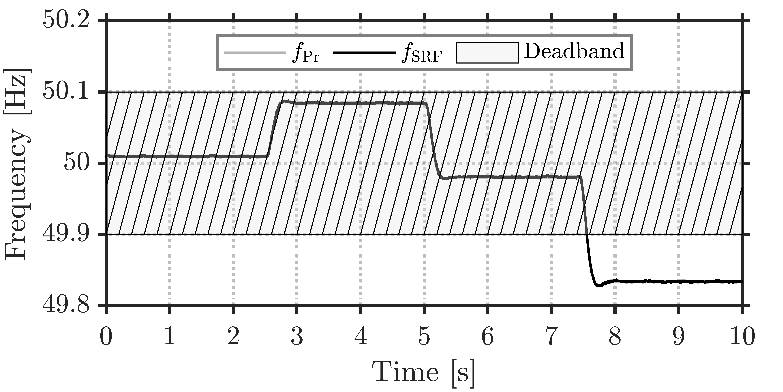}
     \captionsetup{margin={2mm,0mm}}
    \caption{} 
    \label{fig:oper_exp_fr}
    \end{subfigure}
    \vspace{-2mm}
    \caption{System response using the \ac{srf}- and the proposed \ac{pll}.} 
    \label{fig:oper_exp}
    \vspace{-4mm}
\end{figure}

Fig.~\ref{fig:oper_exp} illustrates the system response of the grid-supporting \ac{ibr} employing the proposed and conventional \acp{pll} under various contingencies, including reference power and load changes. The theoretical analysis is validated through the active and reactive power responses and the frequency estimation shown in Figs.~\ref{fig:oper_exp_pw}-\ref{fig:oper_exp_fr}. As expected, the \ac{gfl} converter smoothly tracks the reference setpoint while the system frequency remains within the deadband, thereby nullifying grid-support services. When the frequency exits the $\pm 0.1$~Hz band at $t=7.5$~s due to an abrupt load increase (15~kW, 45~kVar), the power output is immediately adjusted to accommodate the new operating conditions.

\begin{figure}[h!]
    \centering
    \begin{subfigure}{0.239\textwidth}
     \includegraphics[width=\linewidth]{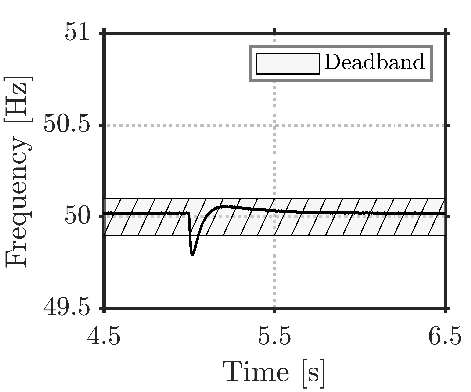}
    \vfill
    \captionsetup{margin={4mm,0mm}}
    \caption{} 
    \label{fig:fr_doubleki}
    %\vspace{2mm}
    \end{subfigure}
    \begin{subfigure}{0.239\textwidth}
        \centering
     \includegraphics[width=1\linewidth]{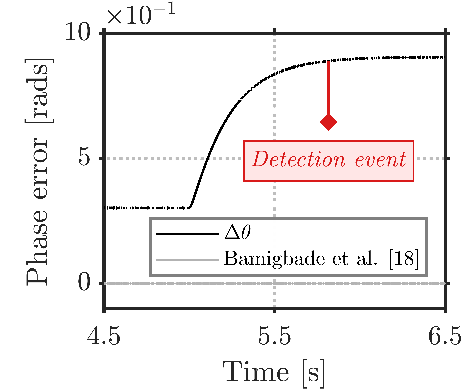}
    %\vfill
    \captionsetup{margin={4mm,0mm}}
    \caption{} 
    \label{fig:atan_doubleki}
    %\vspace{2mm}
    \end{subfigure}\hspace{-4mm}
    \begin{subfigure}{0.239\textwidth}
        \centering
     \includegraphics[width=1\linewidth]{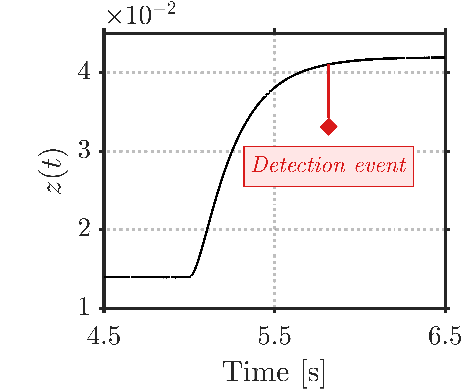}
    %\vfill
    \captionsetup{margin={4mm,0mm}}
    \caption{} 
    \label{fig:dw_doubleki}
    %\vspace{2mm}
    \end{subfigure}
    \begin{subfigure}{0.239\textwidth}
        \centering
     \includegraphics[width=1\linewidth]{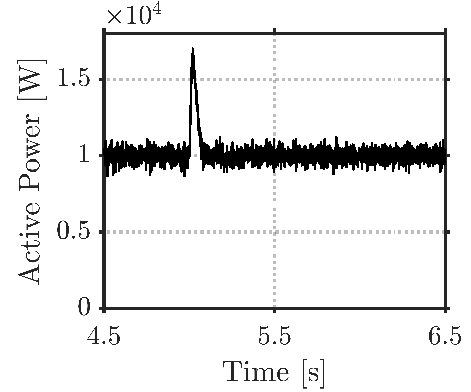}
    %\vfill
    \captionsetup{margin={4mm,0mm}}
    \caption{} 
    \label{fig:power_doubleki}
    %\vspace{2mm}
    \end{subfigure}
    \begin{subfigure}{0.48\textwidth}
     \includegraphics[width=\linewidth]{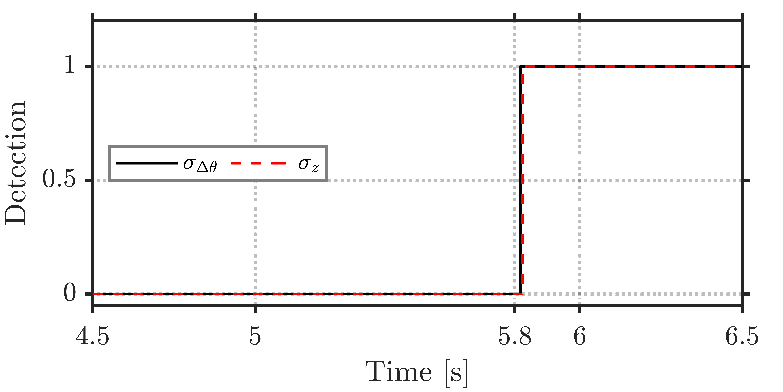}
     \captionsetup{margin={2mm,0mm}}
    \caption{} 
    \label{fig:int_only}
    \end{subfigure}
    \caption{System response after integral-only gain tampering.} 
    \label{fig:firstexp}
    \vspace{-2mm}
\end{figure}

To demonstrate the attack-detection capability of the proposed \ac{pll}, the adopted threat model is applied by actively altering the \ac{pll} gains of the frequency estimator. For comparison, the detector introduced in~\cite{Cyb_attacks}, which enables detection based on the phase estimation error, is also considered. In the first scenario, the adversary sets $k_i^{a} = k_i^{n}/3$ at $t=5$~s. The results are shown in Fig.~\ref{fig:firstexp}. Immediately after the attack, all system states are perturbed, including the active power and estimated frequency depicted in Figs.~\ref{fig:fr_doubleki} and~\ref{fig:power_doubleki}, which subsequently converge to the same equilibrium point. However, the equilibrium points of the internal phase error (Fig.~\ref{fig:atan_doubleki}) and $z$ (Fig.~\ref{fig:dw_doubleki}) are significantly shifted, indicating a gain manipulation. This is also observed in Fig.~\ref{fig:int_only}, where both flags switch to one approximately 0.8~s after the inception of the attack, as dictated by $T_d$. In contrast, the detector of~\cite{Cyb_attacks}, based on the phase estimation error of the \ac{srf} \ac{pll}, fails to detect the abnormality, and the system remains undisturbed during the gain change. This behavior is consistent across additional attack scenarios involving more complex gain-tampering patterns.

\begin{figure}[h!]
    \centering
    \begin{subfigure}{0.24\textwidth}
     \includegraphics[width=\linewidth]{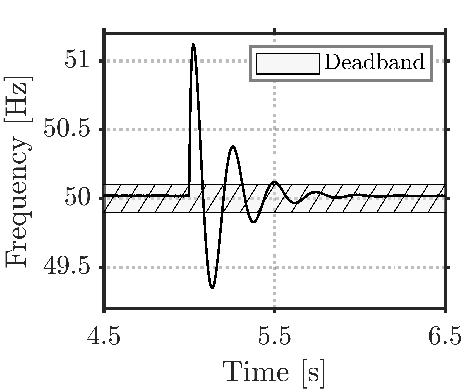}
    \vfill
    \captionsetup{margin={4mm,0mm}}
    \caption{} 
    \label{fig:fr_kihkpl}
    %\vspace{2mm}
    \end{subfigure}
    \begin{subfigure}{0.24\textwidth}
        \centering
     \includegraphics[width=\linewidth]{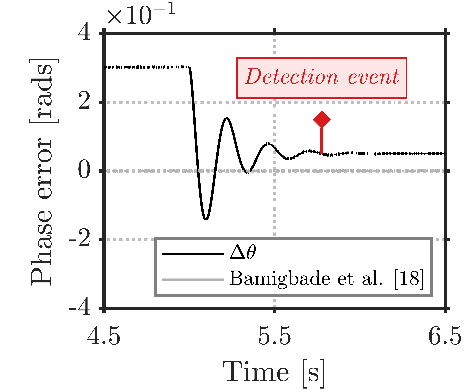}
    %\vfill
    \captionsetup{margin={4mm,0mm}}
    \caption{} 
    \label{fig:atan_kihkpl}
    %\vspace{2mm}
    \end{subfigure}
    \begin{subfigure}{0.24\textwidth}
        \centering
     \includegraphics[width=\linewidth]{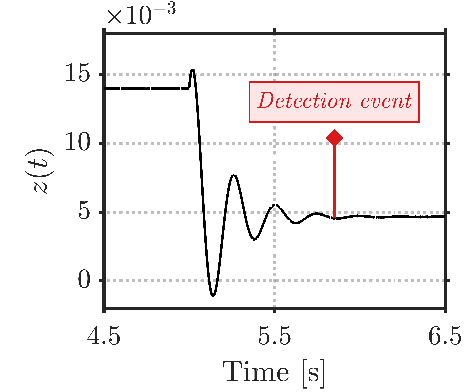}
    %\vfill
    \captionsetup{margin={4mm,0mm}}
    \caption{} 
    \label{fig:dw_kihkpl}
    %\vspace{2mm}
    \end{subfigure}
    \begin{subfigure}{0.24\textwidth}
        \centering
     \includegraphics[width=\linewidth]{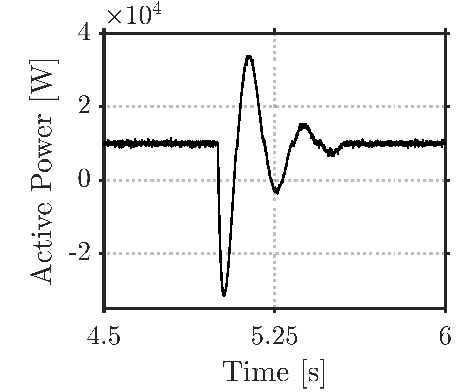}
    %\vfill
    \captionsetup{margin={4mm,0mm}}
    \caption{} 
    \label{fig:power_kihkpl}
    %\vspace{2mm}
    \end{subfigure}
    \caption{System response after proportional-integral gain tampering.} 
    \label{fig:kihkpl}
    \vspace{-6mm}
\end{figure}

\begin{figure}[h!]
    \centering
    \begin{subfigure}{0.24\textwidth}
     \includegraphics[width=\linewidth]{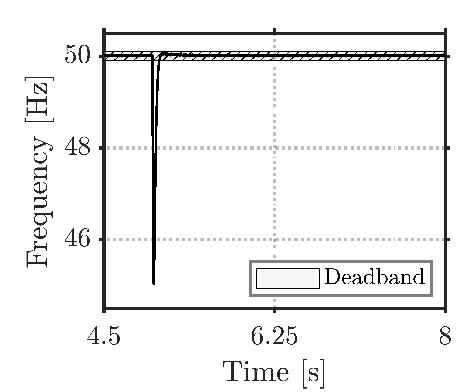}
    \vfill
    \captionsetup{margin={4mm,0mm}}
    \caption{} 
    \label{fig:fr_kpl}
    %\vspace{2mm}
    \end{subfigure}
    \begin{subfigure}{0.24\textwidth}
        \centering
     \includegraphics[width=1\linewidth]{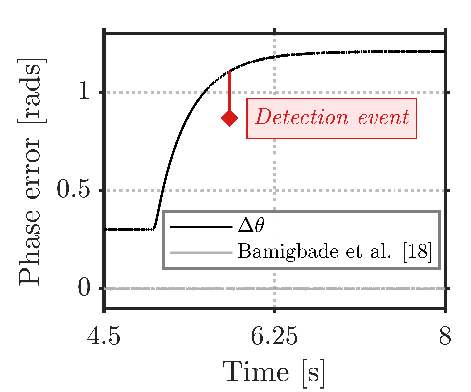}
    %\vfill
    \captionsetup{margin={4mm,0mm}}
    \caption{} 
    \label{fig:atan_kpl}
    %\vspace{2mm}
    \end{subfigure}
    \begin{subfigure}{0.24\textwidth}
        \centering
     \includegraphics[width=1\linewidth]{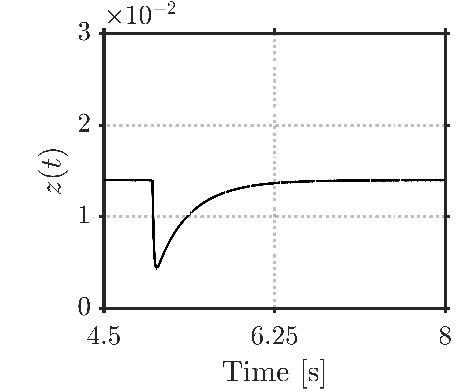}
    %\vfill
    \captionsetup{margin={4mm,0mm}}
    \caption{} 
    \label{fig:dw_kpl}
    %\vspace{2mm}
    \end{subfigure}
    \begin{subfigure}{0.24\textwidth}
        \centering
     \includegraphics[width=1\linewidth]{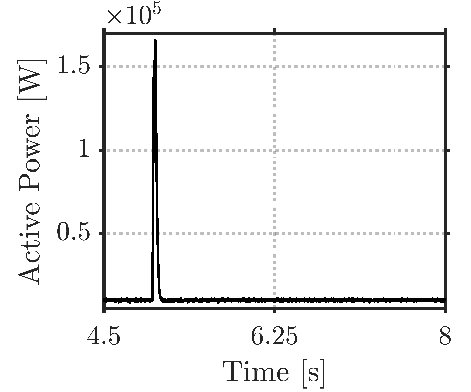}
    %\vfill
    \captionsetup{margin={4mm,0mm}}
    \caption{} 
    \label{fig:power_kpl}
    %\vspace{2mm}
    \end{subfigure}
    \caption{System response after proportional gain tampering.} 
    \label{fig:kpl}
    \vspace{-6mm}
\end{figure}

The attack is also detected when both or only the proportional gain is modified. The system responses for $k_p^{a}=k_{p}^{n}/2$, $k_i^{a}=3k_i^{n}$ and $k_p^{a}=2k_{p}^{n}$, are shown in Figs.~\ref{fig:kihkpl} and~\ref{fig:kpl}, respectively. In both cases, the frequency estimator continues to track the grid frequency (Figs.~\ref{fig:fr_kihkpl} and~\ref{fig:fr_kpl}), while the active power output (Figs.~\ref{fig:power_kihkpl} and~\ref{fig:power_kpl}) is significantly disturbed before converging to the original equilibrium. In contrast, the equilibrium points of $z$ (Figs. \ref{fig:dw_kihkpl} and \ref{fig:dw_kpl}) and the phase error (Figs. \ref{fig:atan_kihkpl} and \ref{fig:atan_kpl}) are notably shifted, enabling attack detection. As in the previous scenario, because the integral action remains active, the detector of~\cite{Cyb_attacks} cannot identify the attack, which remains completely stealthy from its perspective. In practice, the minimum detection delay is upper-bounded by the sum of the settling times of the conditioning filter ($\approx 0.04$~s) and the \ac{pll} ($\approx 0.34$~s), since the two stages respond in series; this yields a conservative estimate of roughly $0.38$~s, the true combined response being somewhat faster. The dwell time $T_d$ can therefore be set from this floor, for fastest response, up to a few seconds; here it is set higher to cover the settling of any \ac{pll} configuration the operator may employ.


\vspace{-2mm}
\subsection{Performance under Unbalanced and Distorted Grid Conditions}
%\vspace{-1mm}
%
This section assesses the proposed detection mechanism under unbalanced and distorted grid conditions. A stiff grid bus is considered, with phases $a$, $b$, and $c$ set to $0.98\,\angle\,{+}0.5^{\circ}$~pu, $1.05\,\angle\,{-}1.5^{\circ}$~pu, and $0.96\,\angle\,{-}0.5^{\circ}$~pu, respectively, introducing a non-negligible negative-sequence component in addition to the nominal positive-sequence voltage. To reject the resulting distortion, the \acp{pll} are equipped with decoupling \ac{qsg} networks \cite{sogi}, and the synchronization \ac{pll} bandwidth is set to 25~Hz given the grid stiffness. Furthermore, the deadband is tightened from $\pm 0.1$~Hz to $\pm 0.05$~Hz to stress-test the detection scheme under the resulting increase in primary-controller activity and noise sensitivity. In this benchmark, a single \ac{pll} performs both synchronization and frequency estimation, as in \cite{Exploiting_Inherent}, with $\omega_o$ obtained by low-pass filtering the \ac{pll} frequency at 5~Hz. Here, $f_{N} = 46~\mathrm{Hz} = \omega_{N}/2\pi$, increasing the difference $|\omega_g - \omega_N|$ and thereby enlarging the equilibrium offsets and the sensitivity to the \ac{pll} gains, which facilitates detection. Now, the detection parameters are set to $\epsilon_{\Delta\theta} = 5\times10^{-3}$, $\epsilon_{z} = 1\times10^{-4}$, and $T_d = T_w = 1$~s. Furthermore, the switching frequency is now reduced to 10~kHz to demonstrate that the proposed detection scheme remains effective even under lower sampling rates, underscoring its real-time applicability and industrial integration. Two scenarios are examined. The first establishes that the proposed and conventional \ac{srf} \acp{pll}, both employing \ac{qsg} networks, remain equivalent and robust under these conditions. The second verifies that the detection scheme reliably exposes \ac{pll} gain tampering despite the negative-sequence component and harmonic distortion.

\begin{figure}[t]
\vspace{-2mm}
    \centering
    \begin{subfigure}{0.48\textwidth}
     \includegraphics[width=\linewidth]{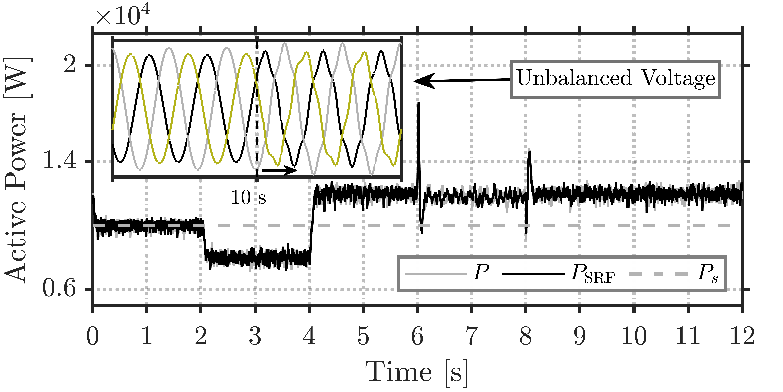}
     \captionsetup{margin={2mm,0mm}}   % shifts label right by 3mm
    \caption{} 
    \label{fig:sogi_exp_pw}
    \end{subfigure}
    \begin{subfigure}{0.48\textwidth}
    \centering
     \includegraphics[width=1\linewidth]{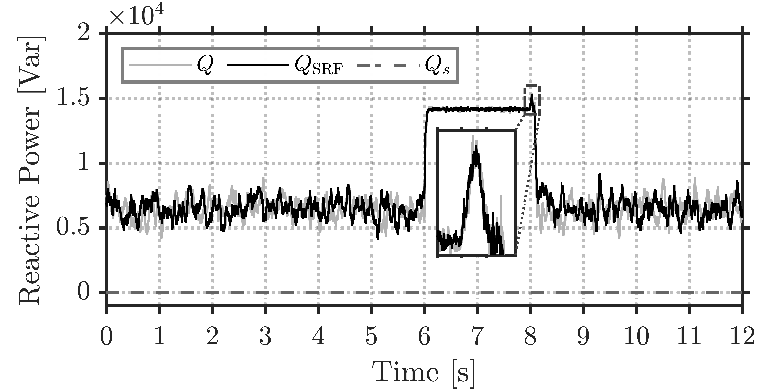}
     \captionsetup{margin={4mm,0mm}}   % shifts label right by 3mm
    \caption{} 
    \label{fig:sogi_exp_rpw}
    \end{subfigure}
    \begin{subfigure}{0.48\textwidth}
     \centering
     \includegraphics[width=1\linewidth]{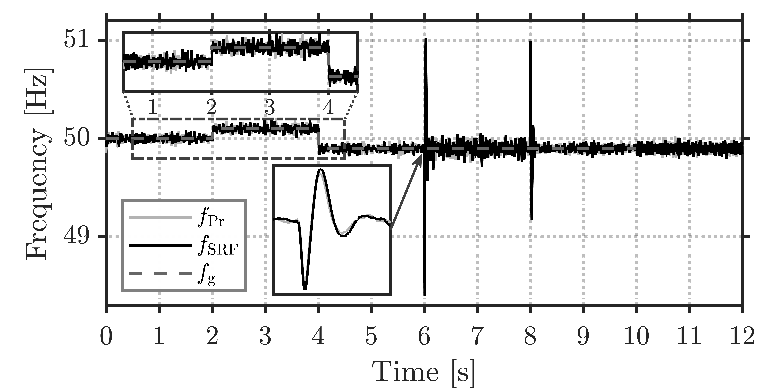}
     \captionsetup{margin={4mm,0mm}}   % shifts label right by 3mm
    \caption{} 
    \label{fig:sogi_exp_fr}
    \end{subfigure}
    \caption{System response using the \acs{srf} and the proposed \acsp{pll} under unbalanced and distorted grid voltage.} 
    \label{fig:sogi_exp}
    \vspace{-6mm}
\end{figure}

The first scenario applies a sequence of disturbances. The grid frequency is stepped to 50.1~Hz at $t=2$~s and to 49.9~Hz at $t=4$~s. A voltage sag is then imposed at $t=6$~s, during which the magnitude of each phase drops to 70~\%, and is cleared at $t=8$~s. Finally, at $t=10$~s, $5^{\rm th}$ and $7^{\rm th}$ harmonics of magnitude 0.06~pu and 0.09~pu are injected, respectively. The system response is shown in Fig.~\ref{fig:sogi_exp}. As in the balanced-grid case, the two implementations perform practically identically. The active power (Fig.~\ref{fig:sogi_exp_pw}) smoothly tracks the outer-loop frequency-support commands following the frequency steps, while the reactive power (Fig.~\ref{fig:sogi_exp_rpw}) adheres to the primary controller, saturating at 14~kVar during the sag at $t=6$~s and recovering to its pre-fault value once the fault clears. The frequency estimates (Fig.~\ref{fig:sogi_exp_fr}) remain virtually identical even through transients, as the zoomed-in plots confirm. Finally, the resilience to higher-order harmonics is evident at $t=10$~s, where both \acp{pll} continue to operate correctly despite the increased harmonic content in the measured signals.

\begin{figure}[h!]
    \centering
    \begin{subfigure}{0.24\textwidth}
        \centering
     \includegraphics[width=\linewidth]{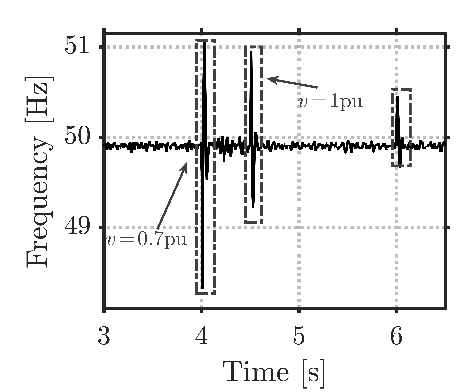}
    \captionsetup{margin={4mm,0mm}}   % shifts label right by 3mm
    \caption{} 
    \label{fig:sogi_fr_kpl}
    \end{subfigure}
    \begin{subfigure}{0.24\textwidth}
        \centering
     \includegraphics[width=\linewidth]{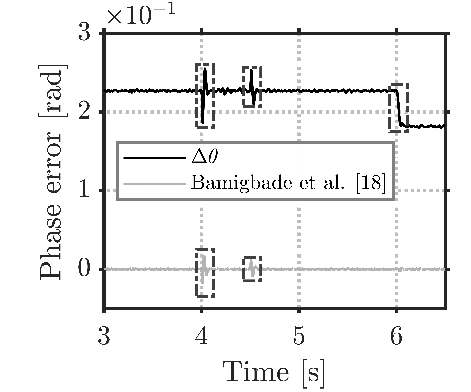}
     \captionsetup{margin={4mm,0mm}}
    \caption{} 
    \label{fig:sogi_atan_kpl}
    \end{subfigure}
    \begin{subfigure}{0.24\textwidth}
        \centering
     \includegraphics[width=\linewidth]{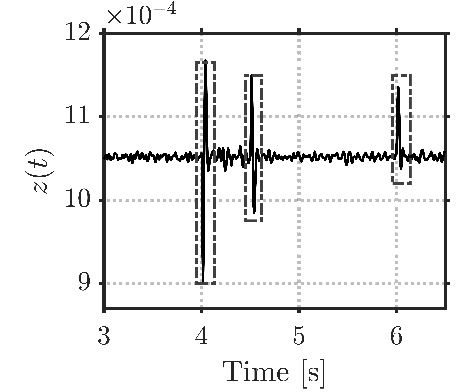}
     \captionsetup{margin={4mm,0mm}}
    \caption{} 
    \label{fig:sogi_dw_kpl}
    \end{subfigure}
    \begin{subfigure}{0.24\textwidth}
        \centering
     \includegraphics[width=\linewidth]{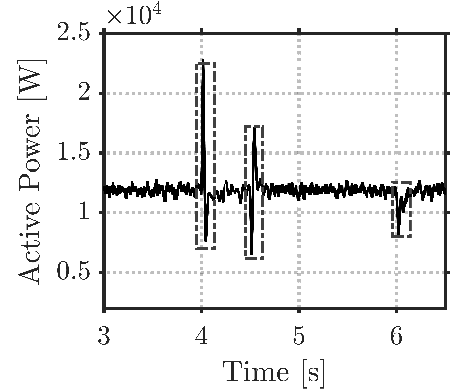}
     \captionsetup{margin={4mm,0mm}}
    \caption{} 
    \label{fig:sogi_power_kpl}
    \end{subfigure}
    \begin{subfigure}{0.48\textwidth}
        \centering
     \includegraphics[width=\linewidth]{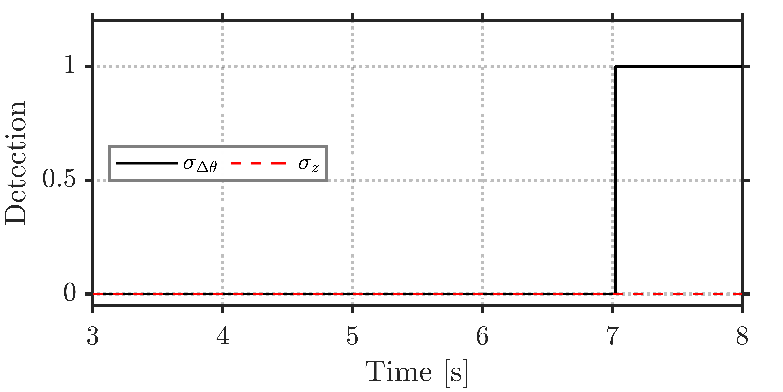}
     \captionsetup{margin={2mm,0mm}}
    \caption{} 
    \label{fig:exp_int_only}
    \end{subfigure}
    \caption{System response after proportional gain tampering under unbalanced and distorted grid voltage.} 
    \label{fig:sogi_kpl}
    \vspace{-6mm}
\end{figure}

The effectiveness of the detection scheme is demonstrated in the second scenario, illustrated in Fig.~\ref{fig:sogi_kpl}, under the same unbalanced grid with constant injection of the $5^{\rm th}$ and $7^{\rm th}$ harmonics of the previous test case. A voltage sag of 0.7~pu occurs at $t=4$~s and is cleared at $t=4.5$~s. The adversary exploits this disturbance as cover, targeting the synchronization \ac{pll} and setting $k_p^{a} = 0.8k_p^{n}$ at $t=6$~s, shortly after the sag clears. Despite the concurrent disturbances, the proposed \ac{pll} operates nominally and continues to track the grid frequency, as shown in Fig.~\ref{fig:sogi_fr_kpl}. Since the \ac{pll} equilibrium points depend only on the grid frequency and on the \ac{pll} gains, neither the voltage sag nor the harmonic distortion can mask the effect of the proportional-gain tampering. The tampering appears as a transient in the power output (Fig.~\ref{fig:sogi_power_kpl}) and in the internal state $z$ (Fig.~\ref{fig:sogi_dw_kpl}), which returns to its pre-attack equilibrium as expected from \eqref{eq:equila} since $z^{\rm s}$ is independent of $k_p$, and, more importantly, as a persistent offset in the equilibrium of the internal phase error (Fig.~\ref{fig:sogi_atan_kpl}), which the plant operator can observe as evidence that the gains have been altered. Evidently, as shown in Fig.~\ref{fig:exp_int_only}, after approximately one second, dictated by $T_d$, the phase-error detector is activated, indicating the tampering, while, as in the balanced-grid case, the detector of \cite{Cyb_attacks} again remains unperturbed and unaware of the compromise. In this case, the  minimum detection delay comprises $\approx 0.04$~s for the conditioning filter and $\approx 0.04$~s for the \ac{pll} settling time, giving a lower bound of $\approx 0.08$~s. As before, this delay is small enough to leave the attack little opportunity to manifest before the operator is alerted; here, however, similarly to the previous scenario, $T_d$ is set higher, to 1~s, to cover the settling times of all plausible gain combinations and potential transients under normal or adverse operating conditions.

\vspace{-2mm}
\section{Conclusion} 
\vspace{-1mm}

This paper demonstrates how an adversary who overcomes local system defenses can manipulate critical control gains of \acp{pll} responsible for synchronizing and controlling grid-supporting \ac{gfl} \acp{ibr}, thereby limiting ancillary services and, in severe cases, destabilizing the underlying \ac{ibr} infrastructure. To characterize this threat, an explicit analysis accounting for all cyber-physical and control aspects of grid-connected converters, including line and grid dynamics, was conducted, revealing the road-map an adversary could follow to compromise \ac{ibr} resources stealthily. The analysis further shows that, while the additional frequency-support loop improves the stability margins and overall small-signal behavior of the converters, poor tuning can provoke severe interactions between control layers and even instability.

To counter this threat, an alternative \ac{pll} implementation based on the \ac{rsu} was introduced. Unlike prior approaches, the proposed scheme preserves the performance of the \ac{srf} \ac{pll}, has low computational complexity and is therefore readily applicable to existing \ac{gfl} infrastructure, and, most importantly, detects both proportional and integral gain tampering by monitoring the internal equilibrium points, which shift whenever either gain is altered. Its effectiveness was validated through extensive \ac{chil} benchmarks under normal, unbalanced, and distorted operating conditions, confirming reliable detection of the otherwise stealthy attack.
\label{sec:conclusion}

%\vspace{-3mm}
%\newpage
\bibliography{references}
\bibliographystyle{IEEEtran}

\end{document}